\documentclass[varenna]{cimento}
\usepackage{graphicx}
\title{Observational issues\ETC\ in radiometric and interferometric
 detection and analysis of the Sunyaev-Zel'dovich effects}
\author{M.~Birkinshaw \atque K.~Lancaster}
\institute{Department of Physics, University of Bristol,
  Tyndall Avenue, Bristol BS8~1TL, U.K.}
\begin{document}

\maketitle

%
%

\section{Single-dish and interferometric observations}\label{sec:obs}

Radiometric observations using single dishes and interferometers are
responsible for most detections of the Sunyaev-Zel'dovich (SZ)
effects. This article discusses the techniques used in measuring the
thermal and kinematic SZ effects in this way, the pitfalls that may
arise, the systematic 
errors in the data, and the resulting uncertainties in the
interpretation of the results. Since these uncertainties limit the
physics return from SZ effect research, some approaches that would
improve this situation are described. Longer reviews of SZ effect
research (Rephaeli 1995; Birkinshaw 1999; Carlstrom, Holder \atque
Reese 2002) may be consulted for additional details.

\subsection{Radiometric quantities}\label{sec:radquan}

In discussing radiometric observations it is conventional to work with
the flux density ($S_\nu$, the energy received per unit time per unit
frequency per unit area) as the measure of the brightness of an
unresolved source. However, the SZ effects of clusters of galaxies are
extended, and so a measure of surface brightness is more
appropriate. Two such quantities are in general use. The flux density
per unit solid angle, $\Sigma_\nu$, is commonly adopted in
describing interferometric observations, where the solid angle is
usually the area of the synthesized beam. The brightness temperature,
$T_{RJ}$, is more usual in single-dish work. 

Flux density per unit solid angle and brightness temperature are
related by 
\begin{equation}
  \label{eq:tbdef}
  \Sigma_\nu = {2 k_B \over \lambda^2} T_{RJ,\nu}
\end{equation}
where $\Sigma_\nu$ and $T_{RJ,\nu}$ are functions of both frequency and 
position. $\lambda$ is the wavelength corresponding to frequency
$\nu$, and $k_B = 1.38 \times 10^{-23} \ \rm J \, K^{-1}$ is the
Boltzmann constant. Eq.~(\ref{eq:tbdef}) can be seen to be based on
the Rayleigh-Jeans approximation for black-body radiation (hence
the use of the RJ suffix). The specific intensity, $I_\nu$, the energy
per unit time per unit frequency per unit area of the detector per
unit solid angle, of black-body radiation is given by the Planck law
\begin{equation}
  \label{eq:planck}
  I_\nu = {2 h\nu^3 \over c^2} \left( \exp\left({h\nu \over k_B T}\right) -
          1\right)^{-1} 
\end{equation}
so that at low frequencies ($h\nu \ll k_B T$) $T_{RJ,\nu} =
T$, the thermodynamic (radiation) temperature. Nevertheless, the
definition of eq.~(\ref{eq:tbdef}) for brightness temperature is used
even at high frequencies, where $T_{RJ,\nu} < T$.

It is important to bear in mind this use of brightness temperature
because observations of primordial fluctuations in the microwave
background radiation (MBR) normally quote results in terms of
fractional changes in the thermodynamic temperature $\Delta T/T$,
rather than $\Delta T_{RJ,\nu}/T_{RJ,\nu}$, or even $\Delta
T_{RJ,\nu}/T$. The relationships between these quantities are
complicated for the thermal SZ effect.

It will also be necessary to discuss the polarization of the SZ
effects. Polarization in radio astronomy is generally described by
the $(I_\nu,Q_\nu,U_\nu,V_\nu)$ Stokes parameters. $I_\nu$
is the specific intensity already described, and measures the
total energy arriving from the source. $(Q_\nu,U_\nu)$ describe that
part of the energy arriving from the source that is linearly
polarized, with a positive value for $Q_\nu$ corresponding to
vertically polarized radiation dominating over the horizontal
polarization. $(I_\nu,Q_\nu,U_\nu)$ together can be used to
calculate the linearly polarized flux density fraction, $\Pi_\nu$, and
its position angle, $\Phi_\nu$,
\begin{eqnletter}
 \label{eq:pol}
 \Pi_\nu  & = & {\left(Q_\nu^2 + U_\nu^2\right)^{1/2} \over I_\nu}
                \label{eq:pol:pi} \\
 \Phi_\nu & = & {1 \over 2} \tan^{-1} \left( {U_\nu \over Q_\nu}
                \right) \quad .
                \label{eq:pol:phi}
\end{eqnletter}
while the circular polarization Stokes parameter, $V_\nu$, will not
be of interest in this article.

It is useful to express the flux density/brightness
temperature relationship for the thermal SZ effect in the usual units
of radio astronomy, as
\begin{equation}
  \label{eq:flux}
  \left( \Delta S_{\rm th, \nu}/{\mu\rm Jy} \right) 
    = 8.16 \,
           \left( \Delta T_{\rm th,0}/{\rm mK} \right) \,
           \left( \nu / {\rm GHz} \right)^2 \,
           \left( \theta  / {\rm arcmin} \right)^2 \,
           f(\nu,T_{\rm gas})
\end{equation}
where $f(\nu,T_{\rm gas})$ is the spectrum of the SZ effect in
brightness temperature terms, normalized to its value at zero
frequency. In the Kompaneets approximation,
\begin{equation}
 \label{eq:komp}
  f(\nu,T_{\rm gas}) = {x^2 e^x \over \left( e^x - 1 \right)^2} \,
                       \left( 2 - {1 \over 2} x \coth {x \over 2}
                       \right) 
\end{equation}
where $x = {h \nu / k_B T}$, and $f$ is independent of the temperature
of the gas in the cluster, $T_{\rm gas}$. A more precise
description of the scattering process (Rephaeli 1995), shows that $f$
is a function of both $\nu$ and $T_{\rm gas}$, and that there are
deviations of the spectrum from eq.~(\ref{eq:komp}) at high temperature.
$\Delta T_{\rm th,0}$ is the zero-frequency
brightness temperature change between a line of sight through a
cluster and an average line of sight that sees only the unscattered
MBR, while $\Delta S_{\rm th,\nu}$ is the flux density difference at
frequency $\nu$ caused by the thermal SZ effect. A rich cluster of
galaxies with $k_B T_{\rm gas} = 5$~keV might have a central thermal
SZ effect $\Delta T_{\rm th,0} = -0.5$~mK at zero frequency. A region
with angular radius $\theta = 0.5$~arcmin in the central part of the
cluster will then appear with a flux density $\Delta S_{\rm th,\nu} =
-0.9$~mJy at 30~GHz (where $\Delta T_{\rm th,\nu} = 
-0.48$~mK), $-6.4$~mJy at 110~GHz ($-0.26$~mK), and $+11.0$~mJy at
350~GHz ($+0.04$~mK) after a null in either flux density or brightness
temperature terms at about 218~GHz.

The corresponding kinematic SZ effect, $\Delta T_{\rm kin,\nu}$ is
smaller than the thermal effect by a factor
\begin{eqnletter}
  {\Delta T_{\rm kin,0} \over \Delta T_{\rm th,0}}
    &=& {1 \over 2} \, {v_{\rm z} \over c} \, 
       \left( {m_e c^2 \over k_B T_{\rm gas}} \right) \\
    &=& 0.085 \, \left( v_{\rm z} / 1000 \ {\rm km \, s^{-1}} \right)
       \, \left( k_B T_{\rm gas} / 10 \ {\rm keV} \right)^{-1}
\end{eqnletter}
at low frequency. Thus if the cluster with $k_B T_{\rm gas} = 5$~keV
and $\Delta T_{\rm th,0} = -0.5$~mK is moving away from the observer
with a peculiar radial velocity $v_{\rm z} = 1000 \ \rm km \, s^{-1}$,
the kinematic effect will produce signals in a region with angular
radius $0.5$~arcmin of $\Delta S_{\rm kin,\nu} = -0.15$~mJy ($\Delta
T_{\rm kin,\nu} = -83 \ \rm \mu K$) at 30~GHz, $-1.6$~mJy ($-63 \ \rm
\mu K$) at 110~GHz, and $-1.7$~mJy ($-7 \ \rm \mu K$) at 350~GHz, with
a maximum flux density effect of $-2.8$~mJy ($-28 \ \rm \mu K$) near
the null of the thermal effect, at about 218~GHz.

While the flux densities for the thermal and kinematic effects are not
small by comparison with the sensitivities achievable by (for example)
the Very Large Array (VLA) or Australia Telescope Compact Array (ATCA)
in a few hours of observing, the relatively large 
angular sizes on which the effects appear cause considerable
difficulties in their detection, as will become apparent later. 

Polarization SZ effects arise from a number of causes, including
multiple inverse-Compton scatterings within clusters and the
transverse or radial peculiar motions of clusters. Recent discussions
of polarization terms have been given by Challinor, Ford \atque Lasenby
(2000), and values 
\begin{equation}
  \Pi_\nu \sim \tau_e 
\end{equation}
where $\tau_e$ is the inverse-Compton scattering optical depth
(normally less than $10^{-2}$) are typical for multiple scatterings in
even the richest clusters of galaxies. Even smaller effects are
obtained from single scatterings of the quadrupolar anisotropy,
or multiple scatterings of the dipole anisotropy, induced by cluster
motions. The detection of such small effects is not yet possible, and 
so their study must await the development of specialized SZ effect
telescopes (Sec.~\ref{sec:iszo}). Nevertheless, such polarization
signals would be of considerable interest since they probe the
kinematics of clusters, and so provide information not otherwise
available about the development of large-scale structure.

As a final element of jargon associated with SZ observing, it is
important to distinguish between the brightness temperature, 
$T_{RJ}$, associated with the properties of the radiation field,
and the antenna temperature, $T_A$, measured by a radio
telescope. $T_A$ and $T_{RJ}$ are related by an efficiency factor
which depends on the ability of the telescope to detect incoming
radiation and the relative sizes and shapes of the telescope beam and
the source on the sky. For our present purposes we will regard the
relationship between the absolute brightnesses of structures seen on
the sky and the detected radio power to be a matter of absolute
calibration embedded in a generic gain factor $G$.

\subsection{Single-dish techniques}\label{sec:singtech}

A single-dish telescope provides an adaptable platform on which to
mount a receiver system. While other arrangements are possible,
receivers and receiver arrays are generally located at either prime or
secondary focus, with the choice of focus depending on the physical
size of the receiver, the size of the focal plane that is to be
occupied by a receiver array, the gain of the telescope, and the
amount of scattered radio power that can be tolerated.

It is useful to begin our analysis of single-dish radiometry by
considering the simplest case, of a single beam of solid angle
$\Delta\Omega$ aligned with the telescope boresight. A receiver behind
such a system will record a power, $P$, which is not only related to
the brightness temperature of the sky within the beam, $T_{\rm sky}$,
but also to any other signals that may appear at the receiver. These
inevitably include some emission from the atmosphere and a parasitic
ground signal, $T_{\rm gnd}$, so that 
\begin{equation}
  P = G ( T_{\rm sky} + T_{\rm atm} + T_{\rm gnd} )
\end{equation}
and no measure of the incoming power, however carefully calibrated,
will be a good measure of the temperature of the sky because of the
unknown contributions of $T_{\rm atm}$ and $T_{\rm gnd}$ to $P$.

The sensitivity of this radiometer system is
\begin{equation}
  \Delta T_{\rm A} = {T_{\rm sys} \over \sqrt{2 \, \Delta\nu \, 
  t}}
  \label{eq:sens}
\end{equation}
where $T_{\rm sys}$ is the system noise temperature, $\Delta\nu$ is
the bandwidth of the receiver, and $t$ is the integration time
used. For systems operating at $\sim 30$~GHz, values of $T_{\rm sys}
\sim 30$~K and $\Delta \nu \sim 1$~GHz are readily obtained, so that
the antenna temperature noise after about 1~hour of integration should
be $\Delta T_{\rm A} \sim 11 \ \rm \mu K$, and the correponding sky noise
(for an antenna with efficiency $0.6$) would be $\Delta T_{\rm sky}
\sim 19 \ \rm \mu K$. While this estimate suggests that sky
temperatures can be measured to high accuracy in a relatively short
time, it is hopelessly optimistic. The sensitivity does not improve as
$t^{-{1 \over 2}}$ over long timescales in measurements with a
single radiometer because of the varying ground and atmosphere
contributions, and because the receiver noise cannot be made ``white''
over such a long period. Furthermore, the interesting astronomical
signal cannot be extracted from the contaminating atmospheric and
ground signals. Thus a different observing technique must be
used. 

The simplest improvement is to move the observing direction between
the point of interest (the target, T) and a reference background
region (R) every few seconds. Subtracting the resulting powers should
provide a measure of the sky temperature difference between the target
and reference region
since most of the atmospheric and ground signals should be slow
functions of time and position. This technique of \textit{position
switching} can provide a good measurement of the temperature
difference $T_{\rm sky,T} - T_{\rm sky,R}$. The angular separation of
the target and reference regions must be sufficiently small that the
atmosphere and ground signals are similar at the two
locations. However, the differencing scheme then means that only
sky structures which differ at the target and reference region can be
detected. Since half the observing time is spent on either
the target or reference region, and then two noisy measurements have
to be subtracted, the maximum sensitivity of the system is a factor
$2$ worse than the estimate in eq.~(\ref{eq:sens}).

A difficulty with this type of observation is that it relies on
conditions being stable during the interval between observing the
target and reference regions. Since large telescopes tend to move
slowly, large changes in the contaminating signals are likely under
any but the best conditions, and smaller changes are expected under
all conditions because of the varying position of the telescope as it
tracks the target across the sky. This problem can be alleviated by
moving the beam's position using an oscillating mirror, or a nutating
subreflector, rather than by moving the entire telescope. Although
this can reduce the position-switching time to a second or less, 
the moving optics almost inevitably induce systematic differences in
the parasitic signals entering the receiver, and there is a mechanical
limit to the rate of position-switching.

An alternative strategy is therefore to observe with two beams,
provided by two separate feeds. The two feeds are placed symmetrically
about the boresight of the telescope, so that the beams have
the same shape on the sky, and the target region is placed in
one beam (the \textit{main} beam, beam~A) while the reference region
lies in the other beam (beam~B). The signals from these two directions
are constantly measured and compared by the receiver system, whose
internal switching between the beams can occur on millisecond
timescales --- sufficiently fast to freeze out atmosphere, ground, and
receiver fluctuations. The resulting measurement of 
\begin{equation}
  \Delta P_{AB} = P_A - P_B
\end{equation}
is averaged over the desired integration time. This
\textit{beam-switching} method of observing is faster than 
position-switching, and reduces the dead time when no data are being
taken. Beam-switching can be highly effective because of its superior
removal of parasitic signals. A problem that remains is that the
target and reference region are being observed with different feeds,
and so different systematic errors on the two sides of the receiver
can lead to systematic errors in the measurement. 

A further improvement is then to combine beam-switching and
position-switching. Now an observing cycle of duration
$t_{\rm cy}$ is broken into three segments
\begin{enumerate}
\item beam~A is off target, and beam~B is pointed at the target with
 the difference signal $\Delta P_{AB}$ integrated over time ${1 \over
 4} t_{\rm int}$ ($s_1 = \int \Delta P_{AB} \, dt$); 
\item beam~A is pointed at the target, and beam~B is off target with
 the difference signal integrated over ${1 \over 2} t_{\rm int}$ ($s_2$);
 and 
\item beam~A is off target, and beam~B is pointed at the target with
 the difference signal integrated over ${1 \over 4} t_{\rm int}$ ($s_3$)
\end{enumerate}
and then the best estimate of the sky brightness difference between
the target and the average brightness of two reference regions offset
to either side of the target is proportional to
\begin{equation}
  s = 2s_2 - s_1 - s_3 \quad . 
\end{equation}
This symmetrical switching pattern is relatively efficient at reducing
the levels of noise induced by time-varying atmosphere and ground
signals, and changing receiver characteristics, since it takes out
linear drifts in the behaviour of the system. Typically the
integration time, $t_{\rm int}$, is $(80 - 90)\%$ of the time taken
for the complete observing cycle, $t_{\rm cy}$, with the lost time
being taken up by moving the telescope.

It is still necessary to design the equipment to reduce non-ideal
behaviour to the maximum extent possible. The receiver should be
designed to have similar gains on the two sides with the
minimum possible signal losses, and maximum symmetry of illumination
of the telescope. The switching scheme should place beams~A and~B
through similar columns of atmosphere, so that switching is generally
performed in azimuth. The ground signal can be equalized if the terrain
near the telescope is flat and unstructured, or carefully
screened, and should be reduced further by designing the feeds to
under-illuminate the antenna. 

\begin{figure}
  \centering
  \includegraphics[width=0.6\linewidth, clip=]{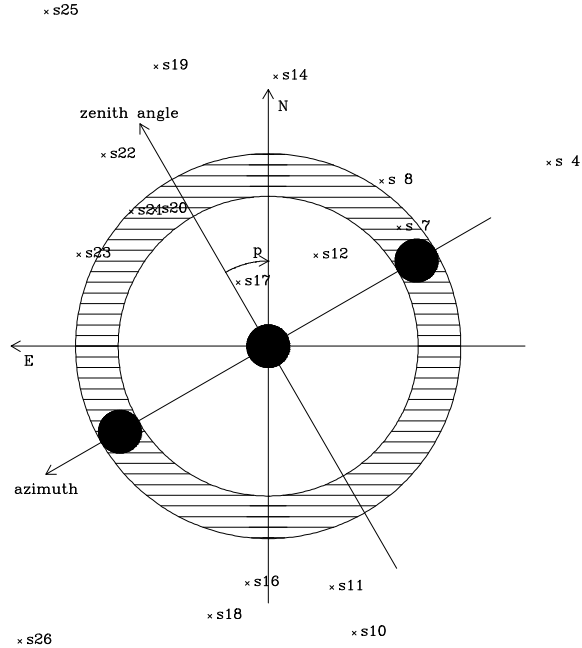}
  \caption{Target position, and the development of reference arcs, for
    OVRO 40-m observations of a point near the centre of
    Abell~665. The current target and reference positions are shown as
    solid circles on a line of constant azimuth. As the sky rotates
    during the observation the reference beams sweep out arcs about
    the target position. For some parallactic angles, $p$, these arcs
    lie near contaminating radio sources: in the configuration shown,
    the NW reference beam lies near sources s7 and s8.}
  \label{fig:arcs}
\end{figure}

Since the observation of an SZ effect may take a number of hours,
spread over a number of days, the positions of the reference beams
rotate on the sky about the target to populate \textit{reference 
arcs} which may extend into a full circle for a circumpolar source
(Fig.~\ref{fig:arcs}). 

As with all astronomical measurements, the observation of an SZ effect
will be subject to confusion from foreground and background
sources. Beam- and position-switched observing has a higher level of
confusion than simple single-beam observing since sources may
lie near both the target and reference beam positions. Fortunately,
switching in azimuth modulates the confusing signal according to the
parallactic angle, $p$. Contaminated data can then be filtered
according to the value of $p$ and removed. With sufficient
instantaneous sensitivity, binning by 
parallactic angle can be used to find and remove even variable
sources, which tend to be a large fraction of the source population at
the high radio frequencies favoured for SZ effect work.

The discussion above has concentrated on two-beam systems, and a
simple position-switching strategy, but this concept can be extended
to receiver arrays and more complicated switching schemes designed to
remove higher-order systematic effects. Additional
levels of switching can also be added: for example, identical
observations may be made on fields leading and trailing the target
field, and the observations can be timed so that the fields
are observed over the same range of hour angle. This can remove a
residual level of environmental systematic error. The receiver can be
rotated, so that the main and reference beams exchange identities, to
remove a level of asymmetry in the data caused by imprecise balance in
the receiver system. Observations can be conducted at different
times of year, so that the position of the Sun and the associated
heating of the telescope and ground change, allowing checks for
the major spillover signals to be performed.

\subsection{Single-dish problems}\label{sec:singprob}

Single-dish observations can potentially provide rapid detections of
SZ effects, particularly if the telescope is equipped with a
radiometer array, but the technique faces three generic problems: of
cluster selection, calibration, and confusion.

\subsubsection{Cluster selection}\label{sec:singsel}

\begin{figure}
  \centering
  \includegraphics[width=0.6\linewidth, clip=]{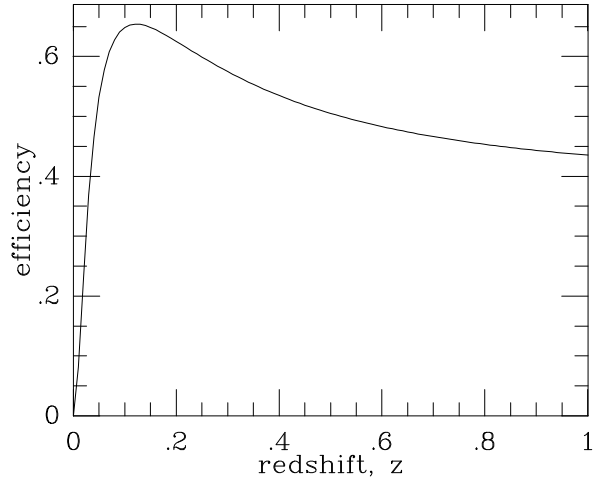}
  \caption{The observing efficiency factor, $\eta$, as a function of
    redshift, for observations of clusters with an isothermal $\beta$
    model atmosphere (eq.~\ref{eq:beta}) with core radius $300 \ \rm
    kpc$ and $\beta = 0.67$ by a telescope with a Gaussian beam of
    108~arcsec FWHM. The decrease of $\eta$ is slow at large redshift,
    and so this observing configuration provides an effective means of 
    detecting clusters at $z > 0.05$ is their intrinsic shape does not
    evolve.}
  \label{fig:zdep}
\end{figure}

While beam- and position-switching have the advantage of removing the
atmosphere and ground signals to good accuracy, they introduce a
selection effect because of the range of angular sizes of SZ effects
that can be observed efficiently. SZ effects of small angular size
fill only a small fraction of the beam, and so produce
only a relatively small signal. SZ effects of large angular size fill
not only the target beam, but also the reference beams, so that there
is little difference in the brightnesses of the sky in the
reference and target beams, so only a small detectable signal. Between
these limits there is some angular size, and hence some redshift, for
which a given telescope and switching system are most sensitive
(Fig.~\ref{fig:zdep}).

The selection imposed by this variation of observing efficiency with
redshift is relatively simple, and the change of efficiency with
redshift is usually slow beyond some redshift of highest sensitivity,
so that this is rarely a major problem.

\subsubsection{Calibration}\label{sec:singcal}

A crucial requirement is that of precise calibration of the brightness
scale. Such absolute measurements are essential if the thermal SZ
effect is to be used to estimate the value of the Hubble constant, or
if the spectrum from a cluster is to be used to separate the thermal
and kinematic effects and so to measure the cluster's peculiar radial
velocity (Sec.~\ref{sec:velocity}).

Calibration of radio telescopes is usually performed by reference to a
set of unresolved radio sources with known flux densities. Absolute
measurements of flux densities are possible for only a few extremely
bright radio sources, and only sparse flux density measurements have
been made. These sources are too bright and too extended to serve as
useful calibrators for single-dish (or interferometric) SZ
observations. Thus it is necessary to transfer these
absolutely-calibrated flux densities to a network of unresolved,
non-variable, secondary calibrators. The interpolation or extrapolation
of these known flux densities to the precise frequency of any given
SZ effect measurement requires an assumption about the shapes of the
sources' spectra. The spectra are generally assumed to be low-order
power-laws in $\log S_\nu - \log\nu$ space, but the sparseness of the
absolute measurements can make the calibrators' flux densities
somewhat uncertain. The combination of this uncertainty
and the difficulties of the original absolute calibration are such
that the flux density scale in normal use may contain systematic
errors at the 5\% level.

An alternative is to calibrate the flux density scale using planets,
again making some assumption about the types of spectra that they
provide, and taking proper account of emission from their atmospheres
and surfaces and the variations in brightness across their disks. This
is not an entirely straightforward process, and includes issues of
polarization effects, the convolution of the brightness distribution
of the planet with the shape of the beam, etc., but provides probably
the best calibration of the flux density scale for a telescope, with
perhaps $3\%$ systematic errors. 

If studies of the spectrum of the SZ effect are to be undertaken, then
the bandpass in which a given observation is made needs to be well
known. In certain frequency ranges the thermal and kinematic SZ
effects have relatively steep spectra, and errors in the
knowledge of the bandpass can lead to errors in
separating the thermal and kinematic components.

It is not enough to make a single calibration of the flux density
sensitivity of an antenna, and then to transfer this calibration to an
internal noise source that can be used for frequent checks of the
system gain. The shapes of large radio telescopes often vary as they
move to different parts of the sky, with a corresponding change in
their gain, and the measurement of this changing gain is also
important if any flux density measurement is to be accurate at the
level of a few per cent.

The beamshape of the telescope must also be well known to interpret
the results being obtained. This may be a complicated process: for a
single dish equipped with a receiver array, the different off-axis
angles of the receiver feeds will mean that each beamshape must be
independently measured. Each of the beams will also have different
polarization characteristics. Both of these properties will change
with telescope elevation. 

Finally, a continuing, real-time calibration of the data is essential.
At cm and mm wavelengths the opacity of the atmosphere changes with
time, and a varying optical depth can change the effective sensitivity
of the system. To some extent the optical depth can be monitored from
the atmospheric brightness (which is reflected in the noise of the
measurements), but at the few per cent level it is essential to obtain
opacity measurements based on sky dips or independent sky monitoring.

\subsubsection{Confusion}\label{sec:singconf}

Observations of any structure on the sky are liable to be confused by
other foreground or background structures, and the problems of
confusion are greater at low angular resolutions and low radio
frequencies than at higher resolutions and higher frequencies. Since
most SZ effects have arcminute angular scales, the effects of
confusion can often be significant and must be taken into account in
interpreting the results.

The most basic confusing signal is that of primordial structure in the
MBR. Such signals have a different spectrum from the thermal SZ
effects, and so in principle this source of confusion could be removed
if sufficiently sensitive data are available at two or more
frequencies, and if the relative calibration of the data is good
enough. In making this separation, the kinematic SZ effect is also
removed since it has the same spectrum as primordial fluctuations in
the MBR. MBR confusion is most important at large angular
scales, but is likely to be $\sim 10 \ \rm \mu K$ at the arcminute
scales which are of most interest for searches for high-redshift
clusters. 

Non-thermal radio sources such as quasars and radio galaxies (and
star-forming objects, particularly at mm wavelengths) also make an
important contribution to the confusion level. The pattern of confusion
can be distorted by the gravitational lensing of the
cluster, and so can cause a variation in the confusing signal with
angle from the cluster centre that mimics the angular structure of the
SZ effects. A large fraction of the background radio source population
is of steep spectrum, and so these problems can be minimized by
observing at short cm wavelengths, or long mm wavelengths. However, a
significant fraction of the radio source population is then variable,
causing the confusion to change with time and adding further
systematic error.

Another possible approach to the reduction of confusion is to detect
the confusing sources using a high-resolution interferometer map, and
then to remove their flux densities from the SZ effect data collected
by the single-dish observations. This approach may run into
difficulties by missing resolved radio emission, but such emission
usually has a steep spectrum and so is minimized by working at high
frequencies. Clusters with strong radio sources are
generally so badly contaminated by their emission that no SZ effect
measurements are possible. A more significant limitation is the need
for frequent interferometric monitoring to deal with 
variable radio sources, which form a significant fraction of the
confusing population at cm wavelengths.

\subsection{Example single-dish results}\label{sec:singex}

Early results from single-dish measurements of the thermal SZ effect
were subject to many unrecognised sources of error, but the
understanding of single-dish data is now good enough that reliable SZ
effect detections of high-temperature X-ray clusters can be achieved
in a few hours per cluster. As a result, many clusters have been
observed by single-dish systems, with random measurement errors $< 100
\ \rm \mu K$, and residual systematic errors (for example from radio
source confusion) of lower levels.

\begin{figure}
  \centering
  \includegraphics[width=0.6\linewidth, clip=]{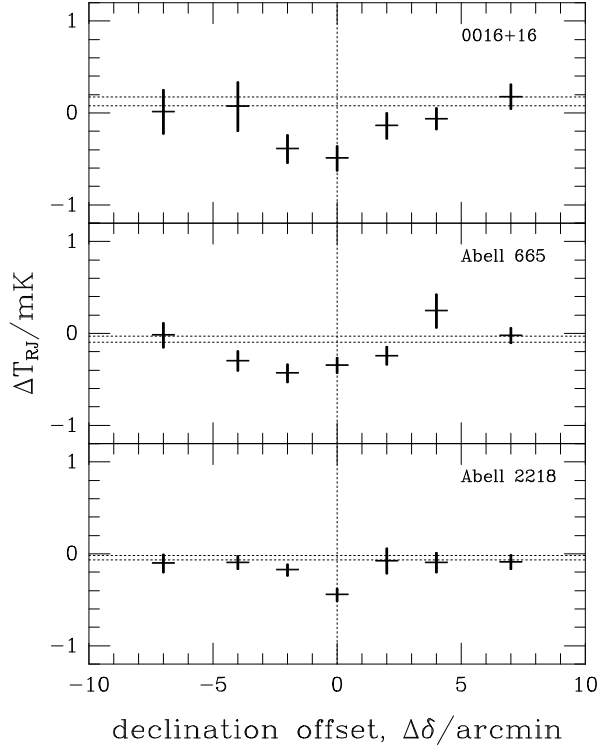}
  \caption{Measurements of changes in the apparent brightness
    temperature of the MBR as a function of declination in the three
    X-ray bright clusters CL~0016+16, Abell~665, and Abell~2218. The
    largest SZ effects are seen near the centres of the clusters, and
    the angular sizes of the effects are consistent with predictions
    based on the X-ray images. The horizontal lines
    mark the range of possible systematic errors in the zero levels on
    the data, and the error bars contain both random and systematic
    components. The brightness temperature scale is subject to a 5\%
    systematic error.}
  \label{fig:threeclusters}
\end{figure}

Some example results (from Birkinshaw 1999) are shown in
Fig.~\ref{fig:threeclusters}. It can be seen that good measurements of
the amplitudes and angular sizes of the SZ effects of
X-ray bright clusters can be made using the beam- and position-switching. 

\subsection{Interferometric techniques}\label{sec:interf}

The technique of interferometry was originally developed with the
intention of achieving high angular resolution, and instruments such
as the VLA are examples of this.  The SZ effect is primarily a large
angular-scale feature on the sky, so high resolution is not of
interest and is, in fact, detrimental in this context.  However,
interferometers offer improvements in the control of systematic
effects compared with single-dish telescopes. First, interferometers
experience a loss of coherence (and thus a loss of sensitivity) away
from the pointing centre, meaning that ground spillover, terrestrial
interference and other spurious signals will be attenuated. Second,
structures on the sky are modulated by a fringe pattern at a
different rate than most contaminating sources.  This allows rejection
of signals from astronomical sources such as the Sun and bright 
planets. Finally, interferometers with a wide range of
baselines allow \textit{simultaneous} observations of confusing, small
angular-scale, possibly-variable radio sources, so that their effects
can be separated from the SZ signal.

\begin{figure}
  \centering
  \includegraphics[width=0.6\linewidth, clip=]{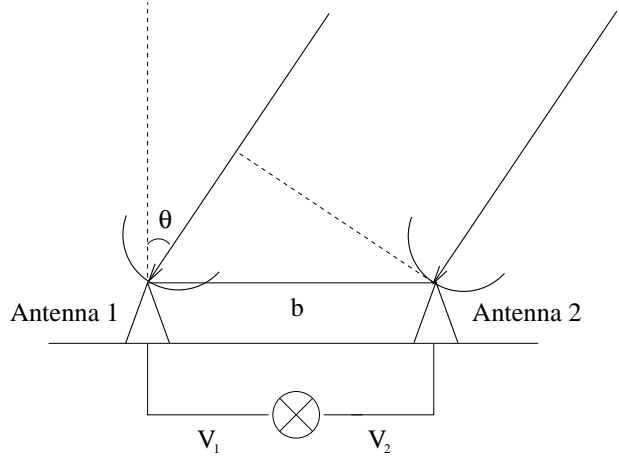}
  \caption{A simple one-dimensional interferometer.  Radiation from
    the source must travel an extra distance $b\sin\theta$ to reach
    antenna 1.} 
  \label{fig:interferometer}
\end{figure}

We can understand the basic principles of interferometry by discussion
of a simple case.  Consider a two-element interferometer, with
antennas separated by a distance $b$, observing a source at an angle
$\theta$, as shown in Fig.~\ref{fig:interferometer}.  Each antenna
receives a signal which produces a time-varying voltage, and the
product of these voltages is measured.  Due to the path difference for
radiation travelling from a distant source to the two antennas, there
will be a phase difference between the received signals given by
\begin{equation}
  \phi = 2 \pi {b \over \lambda} \sin \theta.
    \label{eq:phase}
\end{equation}
The correlated output is then
\begin{eqnletter}
  {\cal A} = V_1 V_2 
           & \propto 
           & \sin(2\pi\nu t) \sin(2\pi\nu t + \phi) \\ 
           & \propto
           & \cos \phi - \cos(4\pi\nu t) + \sin(4 \pi \nu t)\sin\phi
              \quad .
  \label{eq:output}
\end{eqnletter}
In practice the output ${\cal A}$ is integrated over some time
interval so that the rapidly-varying second and third terms average to
zero. If the energy received from the source per unit area is
$S$, and the area of each antenna is $a$, the interferometer
response is
\begin{equation}
  {\cal A} \propto a S \cos \left( 2 \pi {b \over \lambda} \sin\theta
    \right) \quad . 
  \label{eq:response1}
\end{equation}
Phases are usually not measured absolutely, but relative to some
reference direction, $\theta_0$. For a source
offset by a small angle $\Delta \theta$ from $\theta_0$, we have
$\theta = \theta_0 + \Delta \theta$ and (\ref{eq:response1}) becomes  
\begin{eqnletter}
  {\cal A} &\propto& a S \cos \left( 2 \pi {b \over \lambda}
                              \sin(\theta_0 + \Delta \theta)
                            -        2 \pi {b \over \lambda}
                              \sin \theta \right) \\ 
           &\propto& a S \cos \left( 2 \pi {b \over \lambda}
                              \Delta \theta \cos \theta_0 \right)
\end{eqnletter}
since we are dealing with small angles. The correlated output differs
at different antenna separations, so that the angular resolution of this
simple interferometer is proportional to $\lambda / b$.  A more
complex multi-baseline instrument is sensitive to a range of scales
determined by the set of baseline lengths defined by the antenna
locations. The shortest baseline defines the maximum scale which can
be sampled. Sky structures on larger angular scales will not modulate
${\cal A}$ with $\theta_0$ (and hence with time), and so will not
produce a detected signal.

\begin{figure}
  \centering
  \includegraphics[width=0.6\linewidth, clip=]{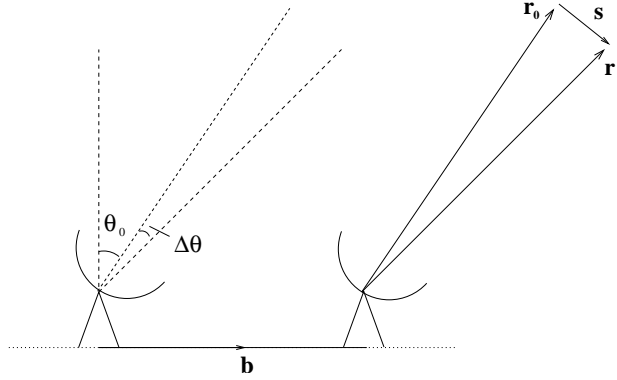}
  \caption{The same simple interferometer as
    Fig.~\ref{fig:interferometer}, where the field centre is 
    specified by $\textbf{r}_0$ and the source position is
    $\textbf{r}$. $\theta_0$ and $\Delta \theta$ are the
    angles discussed in the text.}
  \label{fig:interferometer2}
\end{figure}

The interferometer response can be expressed more generally --- we
consider the main points here, but a full treatment is given in
Thompson, Moran and Swenson (1986).  If we now specify the source
position by a vector $\textbf{r}$ (see Fig.~\ref{fig:interferometer2})
and the baseline by the vector $\textbf{b}$, the phase
difference from (\ref{eq:phase}) can be written
$\phi=(2\pi/\lambda)\textbf{b}.\textbf{r}$. The reference direction
may be specified by a vector $\textbf{r}_0$, so that
$\textbf{r}=\textbf{r}_0+\textbf{s}$, where $\textbf{s}$ describes the
shift between the two.  After some manipulation, the response to all
sources within the solid angle $\Omega$ becomes
\begin{equation}
  {\cal A} \propto \int e^{-2\pi i {\textbf{b}.\textbf{s} \over \lambda}} \, 
           d\Omega \quad . 
\end{equation}

It is conventional to specify the baseline vector $\textbf{b}$
in terms of right-handed coordinates $(u,v,w)$, where $w$ is in
the direction of the source, $u$ and $v$ point East and North
respectively as seen from the source position, and distances 
are measured in wavelengths.  Additionally, the position of the source
on the sky is usually described in terms of co-ordinates
$(l,m,n)$.  We see that
$\textbf{b}.\textbf{r}_0 = w \lambda $, and
$\textbf{b}.\textbf{r}   = (ul+vm+wn) \lambda$, thus
$\textbf{b}.\textbf{s}   = \left( ul+vm+w(n-1) \right) \lambda$.
Making the substitution $d\Omega = (dl\,dm)/n$, we find
\begin{equation}
  {\cal A} \propto \int dl \int dm \, a(l,m) \, I(l,m) \, 
           \frac{e^{-2\pi i \left( u l + v m + w (n-1) \right)}}{n}
  \label{eq:fourier}
\end{equation}
where $a(l,m)$ is the effective total area of the antennas in the
direction $(l,m)$ and $I(l,m)$ is the brightness distribution on the
sky. $n=(1-l^2-m^2)^{1/2} \approx 1$ for small angles, simplifying
the Fourier inversion required in eq.~(\ref{eq:fourier}) to produce a
sky map of $I(l,m)$.  A map made from interferometer data contains
structures which are modulated by the \emph{synthesized beam}. This is
given by the Fourier transform of the telescope aperture, which is
eq~(\ref{eq:fourier}) above with the sky brightness replaced by a
two-dimensional $\delta$~function.

\subsection{Interferometric problems}\label{sec:intprob}

Despite the many advantages in using interferometric techniques there
are numerous complications. Some, like confusion, are of similar type
to those affecting single-dish observations, but there are also new
problems related to bandwidth, integration time and temperature
sensitivity. 

\subsubsection{Confusion}\label{sec:intconf}

\begin{figure}
  \centering
  \includegraphics[width=0.7\linewidth, clip=]{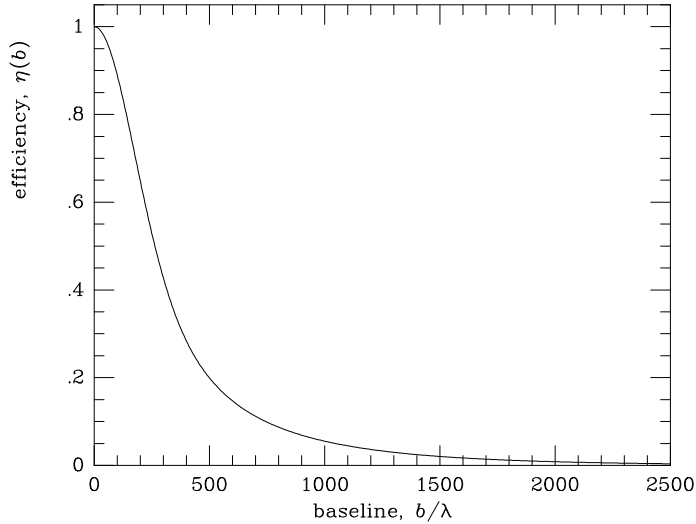}
  \caption{Predicted VLA efficiency as a function of baseline for a
    5-GHz observation of Abell~665. The shortest baseline available on
    the VLA is $\approx 800 \lambda$, so that most of the SZ effect
    signal is resolved out by this long-baseline instrument. Since
    there is no SZ signal on the longest baselines, data on these
    baselines can be used to obtain accurate flux densities of
    contaminating radio sources. These flux densities can then be
    subtracted from the short-baseline data to improve the
    detectability of the SZ effects.} 
  \label{fig:profile}
\end{figure}

Radio source and MBR confusion are difficulties for interferometers
just as for single dishes, but interferometric data implicitly
contain the possibility of filtering out the confusion from small
angular scale radio sources. Short interferometer baselines are
sensitive to the SZ effects and to large angular-scale structures in
the MBR (Fig.~\ref{fig:profile}) as well as to all nearby radio
sources. However features with large angular scale
are resolved out on long baselines, and only radio emission from small
angular scales 
produces detectable signal. Thus the confusing radio sources of small
angular size can be identified using the longest interferometer
baselines, which are insensitive to the SZ effects or MBR structures.
The effects of these sources can then be removed on the short
baselines by a simple subtraction, allowing SZ effects and other MBR
structures to be mapped with much reduced radio source confusion.
This is particularly valuable since the \emph{simultaneous}
observations of SZ effects and radio sources allows the effects of
variable sources to be removed.

This procedure relies on a strong separation between the angular
scales of emission from radio sources and the SZ effects, and
therefore fails for clusters containing bright extended radio emission
(such as radio halo sources). In these cases, the radio emission can
be reduced in importance by working at high frequencies (since it is
generally of steep spectrum), or by using multi-frequency observations
to effect a spectral separation.

\subsubsection{Calibration}\label{sec:intcal}

Calibration issues for interferometers are effectively the same as
those for single dishes, with two caveats. First, interferometers
present the additional issue of phase calibration. The electrical
length of each baseline will vary slightly from day to day due
to instrumental instability (e.g. with ambient temperature), and it is
essential to calibrate this in order to retain spatial information in
the data.  This can be implemented by observation of a bright radio
source, and comparison of the fringe rate observed with that
calculated from a model situation.  Phase corrections can then be
applied to the data. 

Second, it is simplest to calibrate using radio sources that are
unresolved on even the longest baselines, thus reducing the number of
suitable calibrators compared to those available for single dish
work. MBR interferometers have relatively short baselines, but an
additional problem is that their collecting areas are often relatively
small, and consequently their flux sensitivity relatively low.  For
example, the VSA (see Sec.~\ref{sec:intex}) partially resolves the
supernova remnants Cas~A and Tau~A (angular sizes $\sim 5$~arcmin,
flux densities $\approx 180$ and $350$~mJy at 33~GHz) on its longest
baselines, yet the powerful radio galaxy Cyg-A ($\sim 35$~mJy at
33~GHz) is not quite bright enough for accurate phase calibration.

\subsubsection{Bandwidth}\label{sec:intband}

In practice an interferometer does not observe at a single frequency
$\nu$, but rather over a range $\Delta \nu$ about some central
frequency $\nu_0$. This means that the values of $(u,v,w)$, which are
measured in wavelengths, change across the passband, reducing the peak
signal by a factor
\begin{equation}
 \exp\left(-2\left(\pi \frac{\Delta \nu}{\nu_0} \frac{b}{\lambda}
   \right)^2 \right) \quad . 
\end{equation}
This limits the field size, resolution or bandwidth $\Delta \nu$,
which directly affects the sensitivity. To avoid this problem, an
interferometer may split the band $\Delta \nu$ into several sub-bands,
and deal with each sub-band separately.

\begin{figure}
  \centering
  \includegraphics[width=0.5\linewidth, clip=]{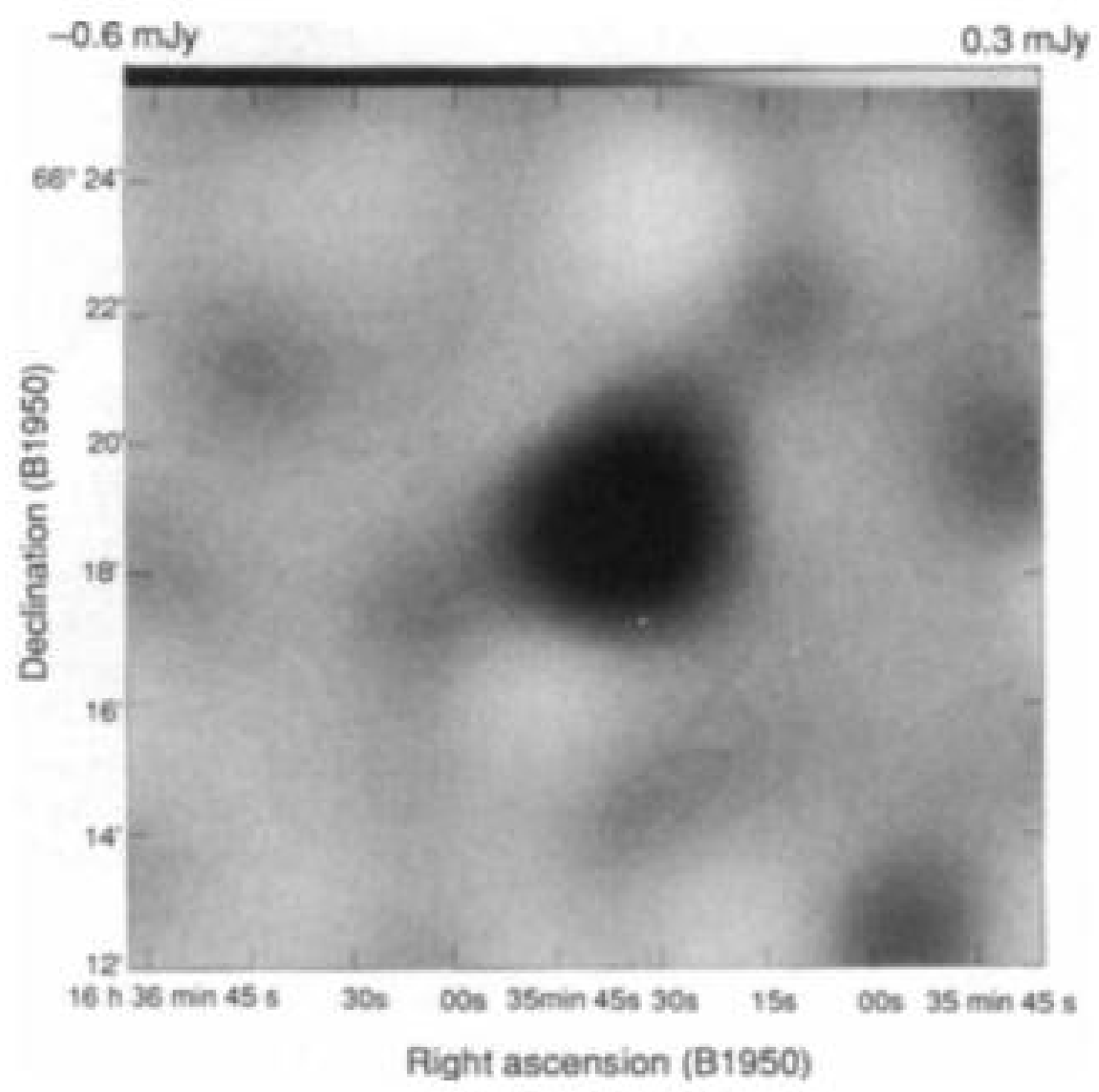}
  \includegraphics[width=0.9\linewidth, clip=]{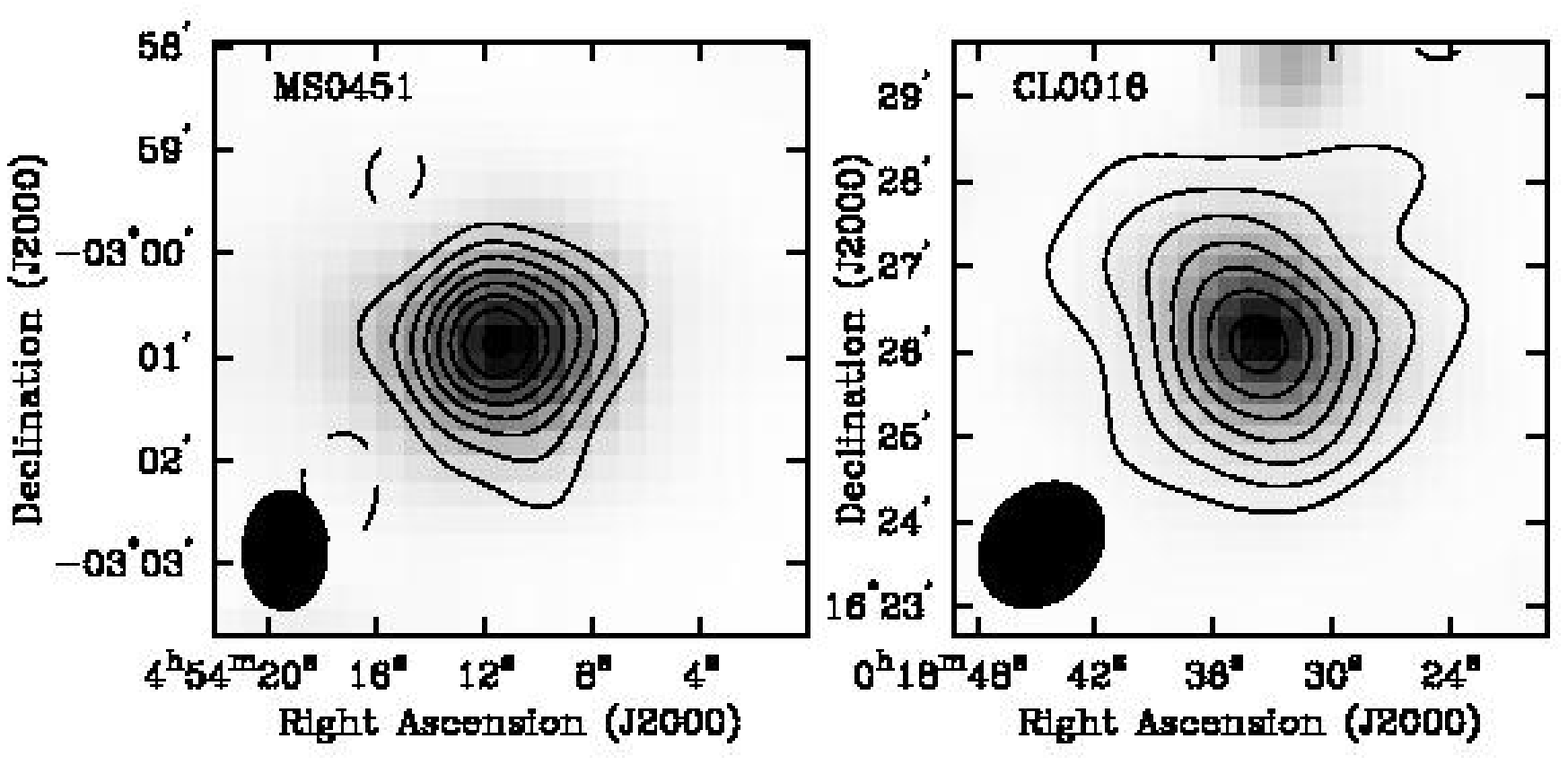}
  \caption{Top: This Ryle telescope observation of Abell 2218 was the
    first interferometric detection of the SZ effect.  Below: The OVRO
    and BIMA arrays have also observed many clusters in SZ, including
    these high significance detections of MS~0451-0354 and CL~0016+16.}
  \label{fig:Ryle-SZ}
  \label{fig:OVRO_BIMA-SZ}
\end{figure}

\subsubsection{Time constant}\label{sec:inttime}

Typically, an interferometer will integrate over a few seconds per
measurement. This causes off-axis sources to be smeared in arcs in
the image plane, again reducing the peak signal.  In order to avoid
loss of precision, the integration time 
\begin{equation}
  \Delta t \ll { {\rm 3 \ hours} \over {\rm number \ of \  synthesized
   \ beams \ across \ field \ of \ interest}} \quad .
\end{equation}
This time-constant smearing is only a small effect for CMB
interferometers, as their short baselines produce relatively large
synthesized beams.

\begin{figure}
  \centering
  \includegraphics[width=0.6\linewidth]{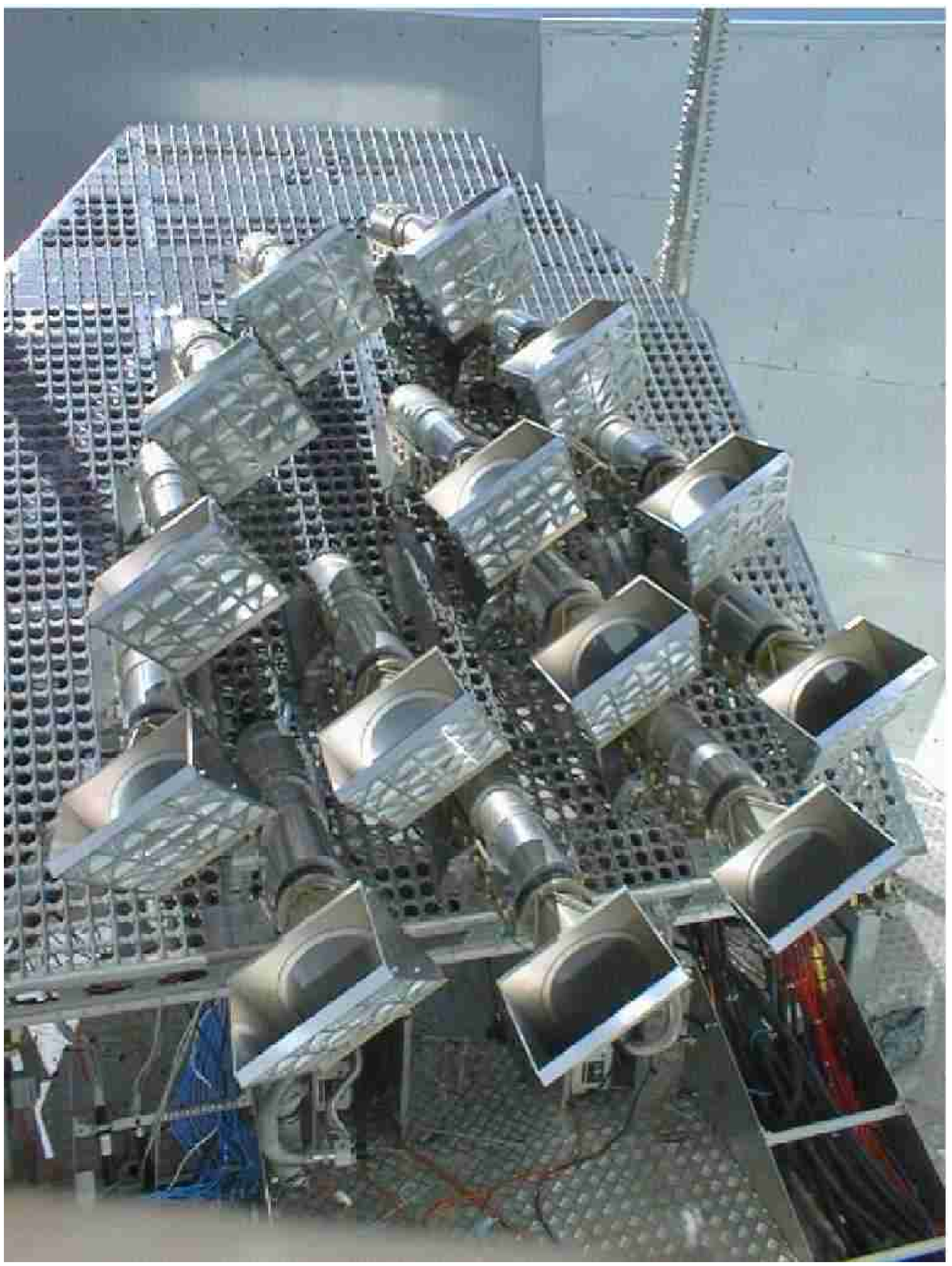}
  \caption{The VSA is a dedicated MBR interferometer located at the
    Observatorio del Teide, Tenerife. Each aperture is about 40~cm in
    diameter and the table is approximately 3~m across.} 
  \label{fig:vsa}
\end{figure}

\subsubsection{Temperature sensitivity}\label{sec:intsel}

The temperature sensitivity of an interferometer is given by
\begin{equation}
  \Delta T_A \propto \frac{T_{\mathrm{sys}}} {\sqrt{\Delta\nu \,
    t_{\rm int} \, N_{\mathrm{corr}}}}
    \frac{1}{\Omega_{\mathrm{synth}}} 
\end{equation}
(compare eq.~\ref{eq:sens}) where $N_{\mathrm{corr}}$ is the number of
antenna-antenna correlations used in making the synthesized beam of
solid angle $\Omega_{\mathrm{synth}}$. This equation shows that
sensitivity increases with increased bandwidth and observing time in
the same manner as for single-dish observing ($t_{\rm int}$
is made up of many measurements, to eliminate time constant
smearing), and also increases with the collecting area of the
telescope, as described by the number of antennas. However, for an
extended source of angular size $\theta_s$, only antennas separated by
baselines less than $\lambda/\theta_s$ contribute to the sensitivity. 
Longer baselines resolve out the signal, as shown in
Fig.~\ref{fig:profile}. The region of high efficiency on this 
normalized \textit{visibility function} becomes narrower for sources
of greater angular extent. As most interferometers are designed for
high point source sensitivity, they tend to contain few short
baselines.  This severely limits sensitivity to the SZ effects, which 
subtend large (arcmin or greater) angular scales on the sky.  One way
to reduce this problem is to to observe at longer wavelengths, however
confusion from radio sources soon becomes a limiting factor. 

Most SZ observations to date have been made using ``normal''
interferometers, i.e., those designed to achieve high resolution rather
than for this purpose.  For example, the Berkeley-Illinois-Maryland
Array (BIMA), which operates primarily at mm wavelengths, has been
used to good effect in its most compact configuration and at cm
wavelengths.  Since few BIMA baselines were short enough to detect the
SZ signal, long integration times were required to achieve appropriate
sensitivity.

\begin{figure}
  \centering
  \includegraphics[width=0.8\linewidth]{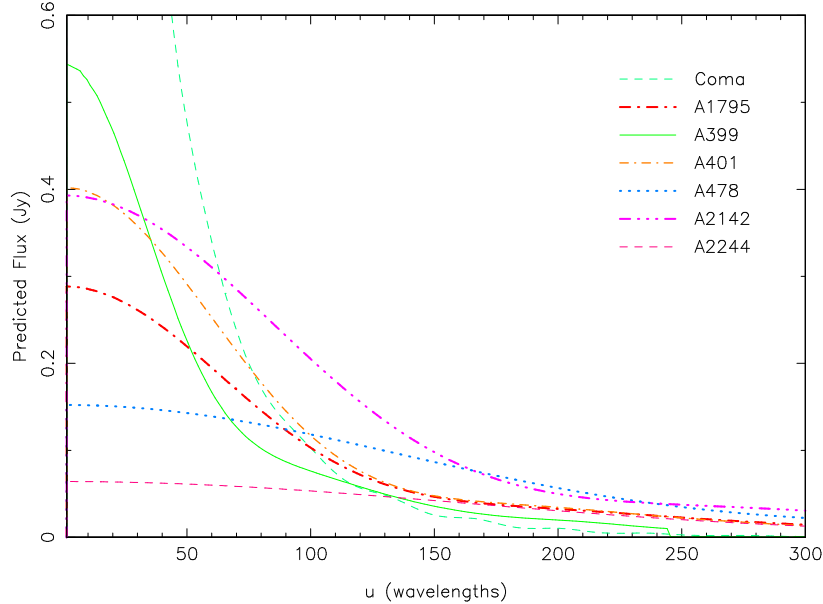}
  \caption{Predicted visibilities for VSA measurements of a sample of
    seven X-ray selected clusters at $z<0.1$.  The shortest VSA
    baseline is $\approx 40\lambda$.}
  \label{fig:vsa-profiles}
\end{figure}

\subsubsection{Cross-talk}\label{sec:intxtalk}

A problem that arises in using compact interferometer configurations is
that of antenna-antenna cross-talk, since microwave signals radiated
from one antenna can leak into adjacent antennas. Such cross-talk
signals can easily dominate the signals expected from SZ effects on
the sky. A reduction of this cross-talk is often possible by using the
different rates of modulation of the cross-talk and sky signals,
at the cost of some loss of data and increased noise.

\begin{figure}
  \centering
  \includegraphics[width=0.3\linewidth, clip=]{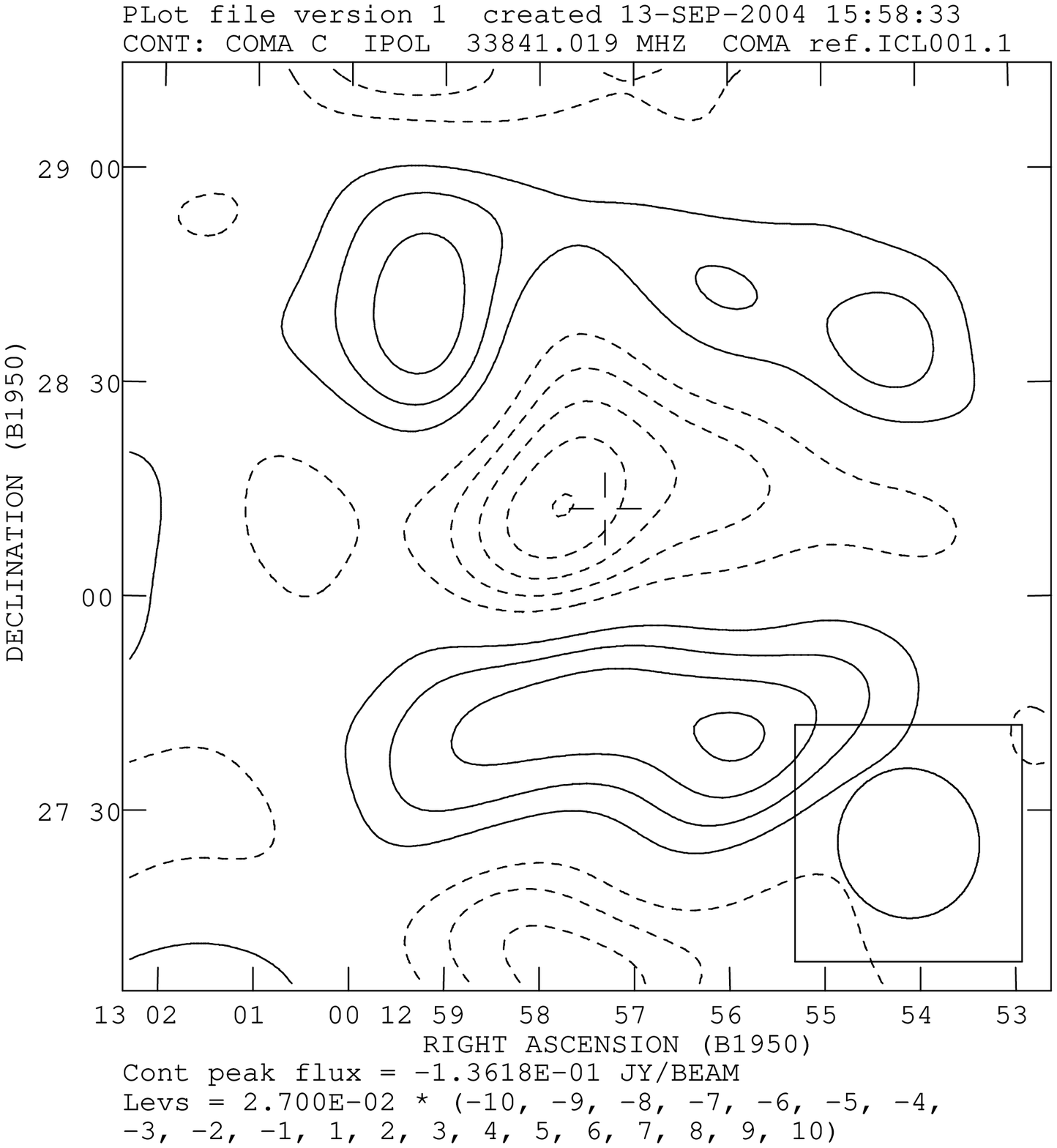}
  \includegraphics[width=0.3\linewidth, clip=]{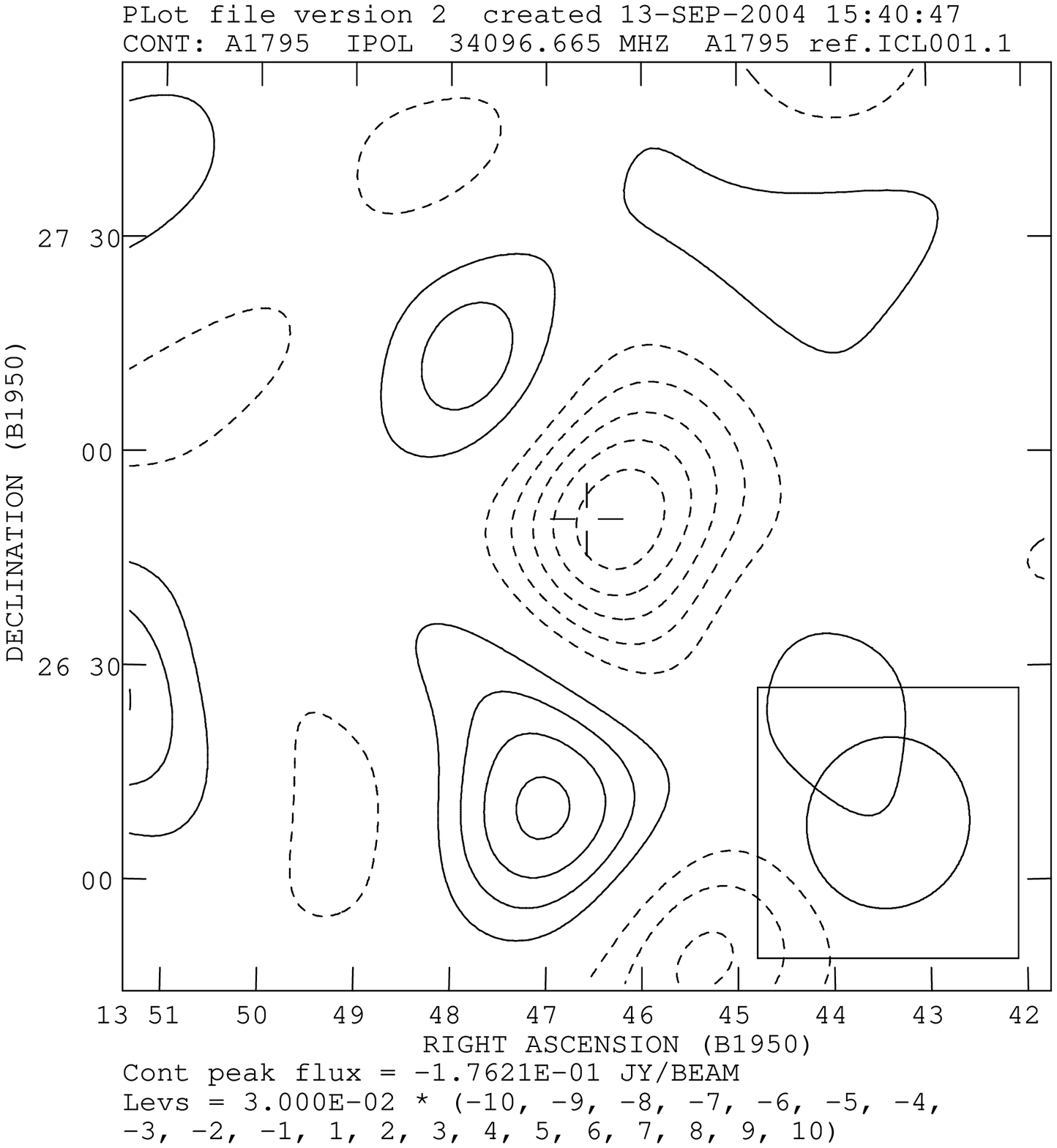}
  \includegraphics[width=0.3\linewidth, clip=]{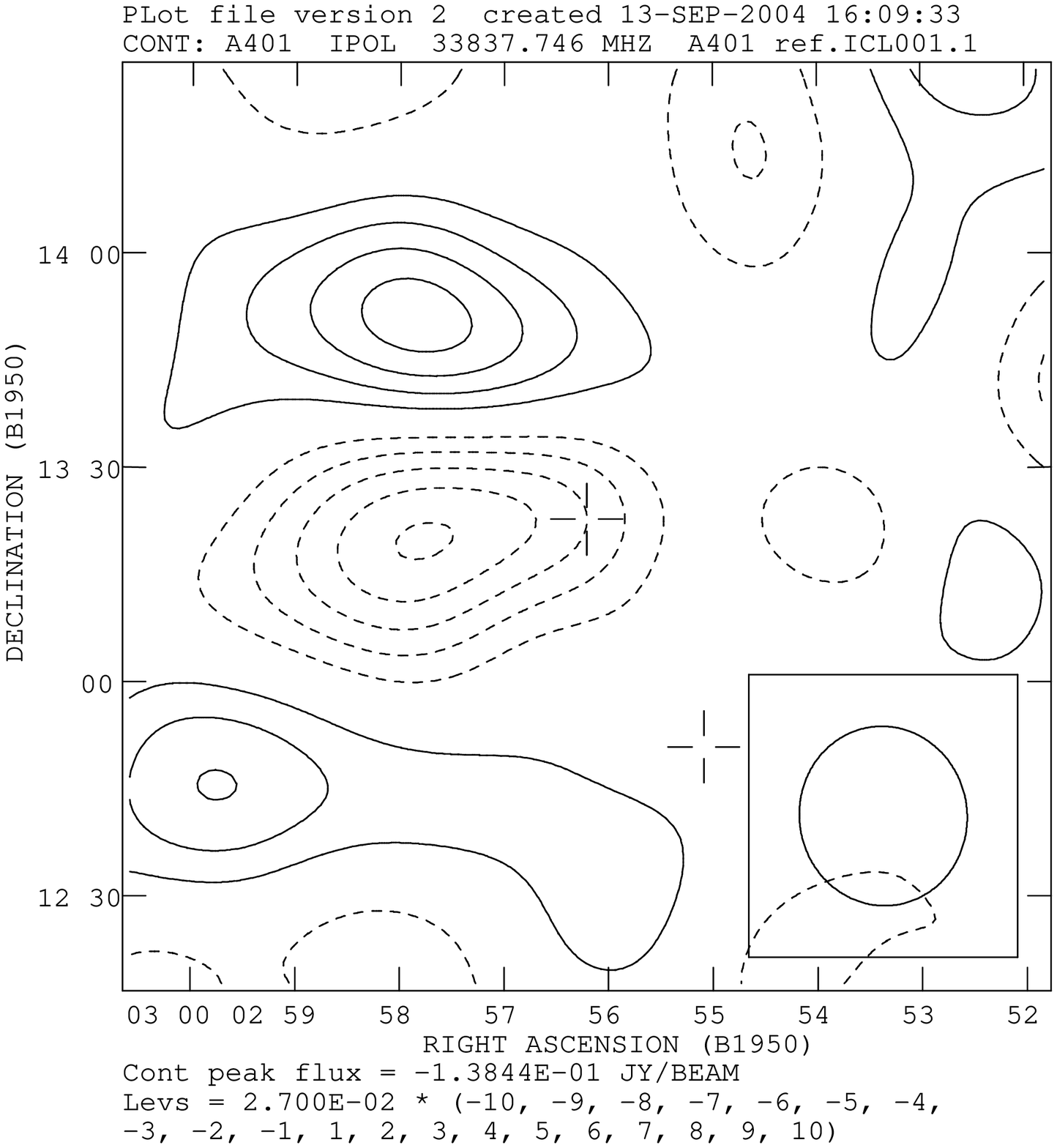}
  \includegraphics[width=0.3\linewidth, clip=]{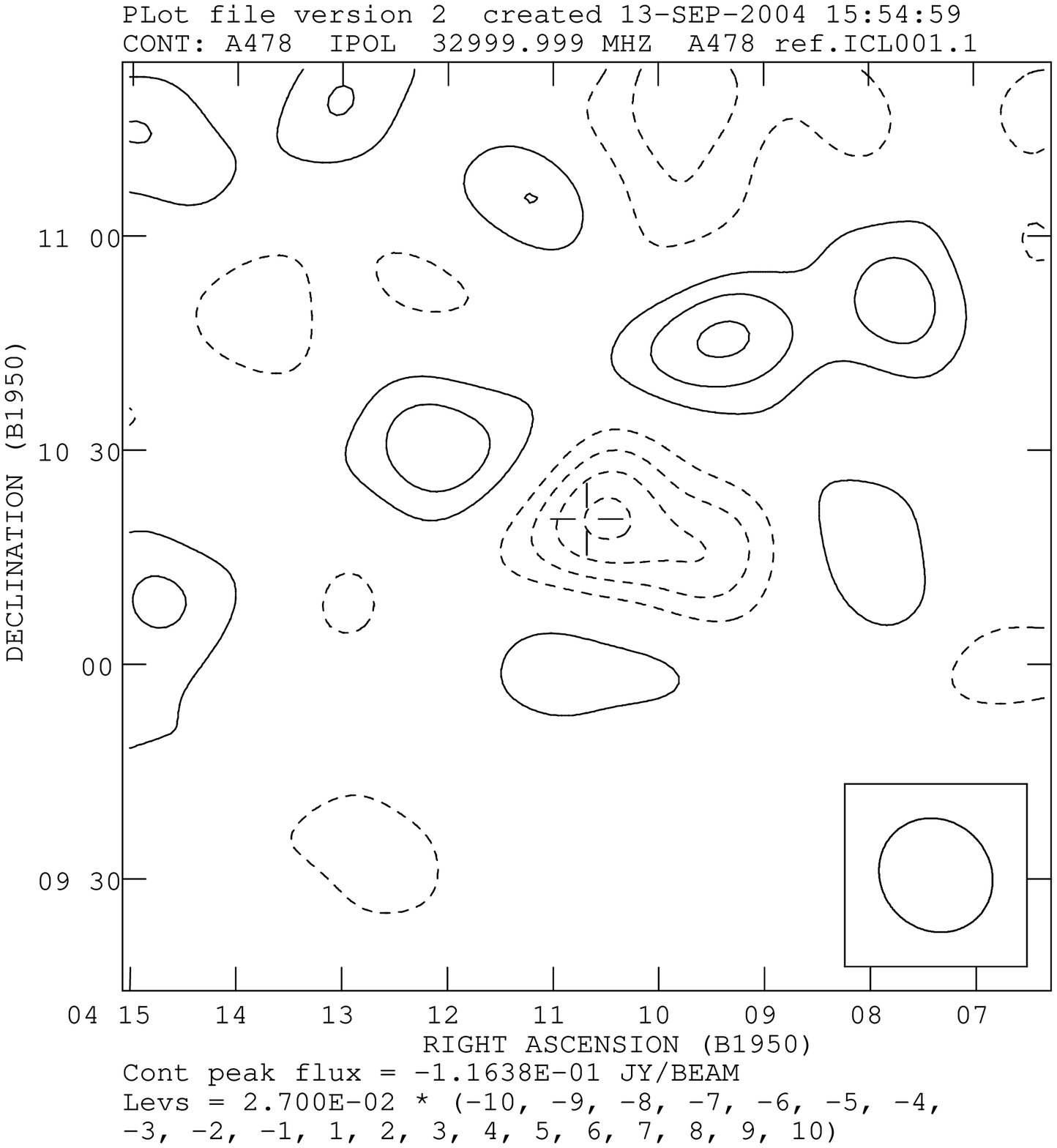}
  \includegraphics[width=0.3\linewidth, clip=]{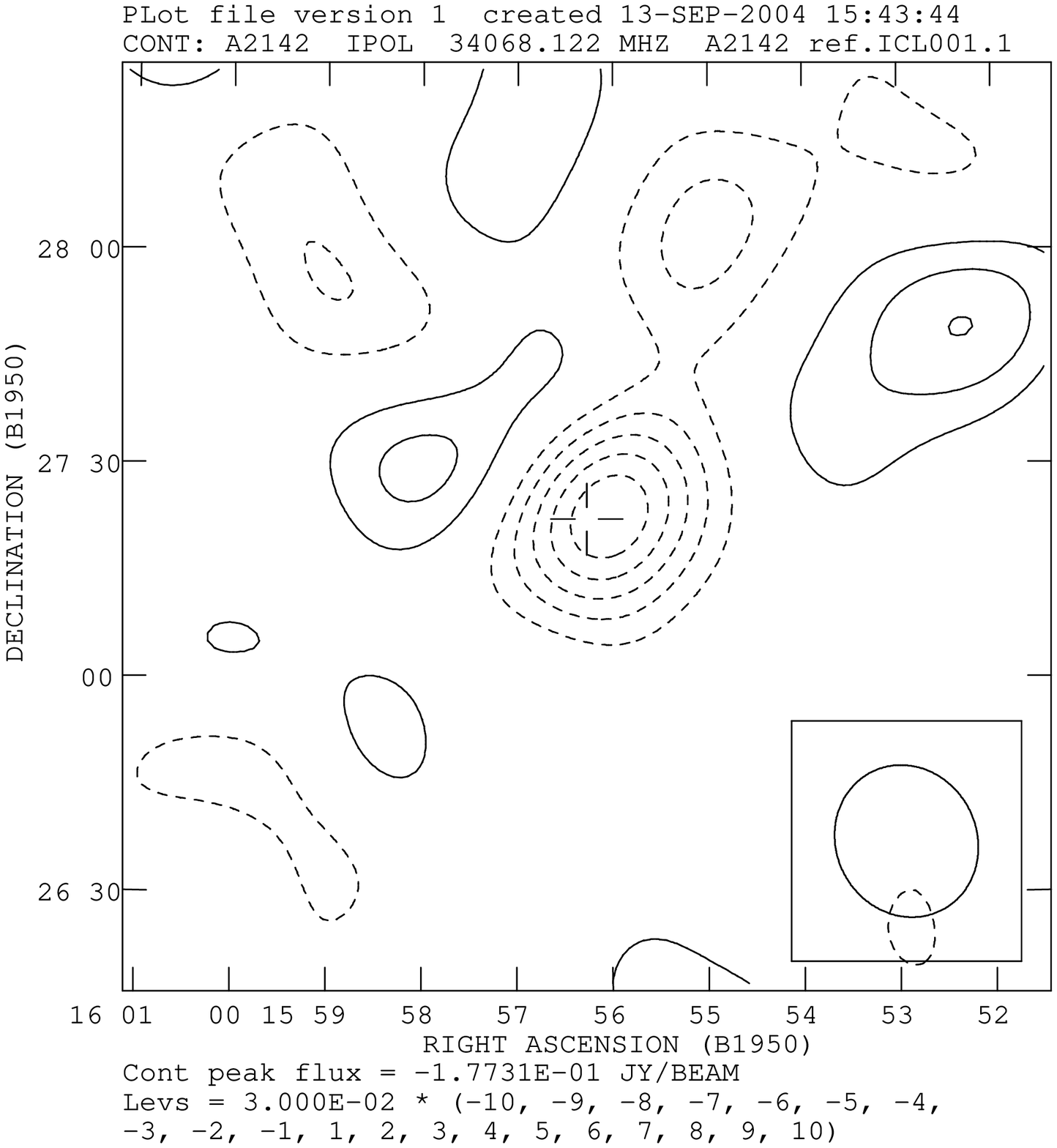}
  \includegraphics[width=0.3\linewidth, clip=]{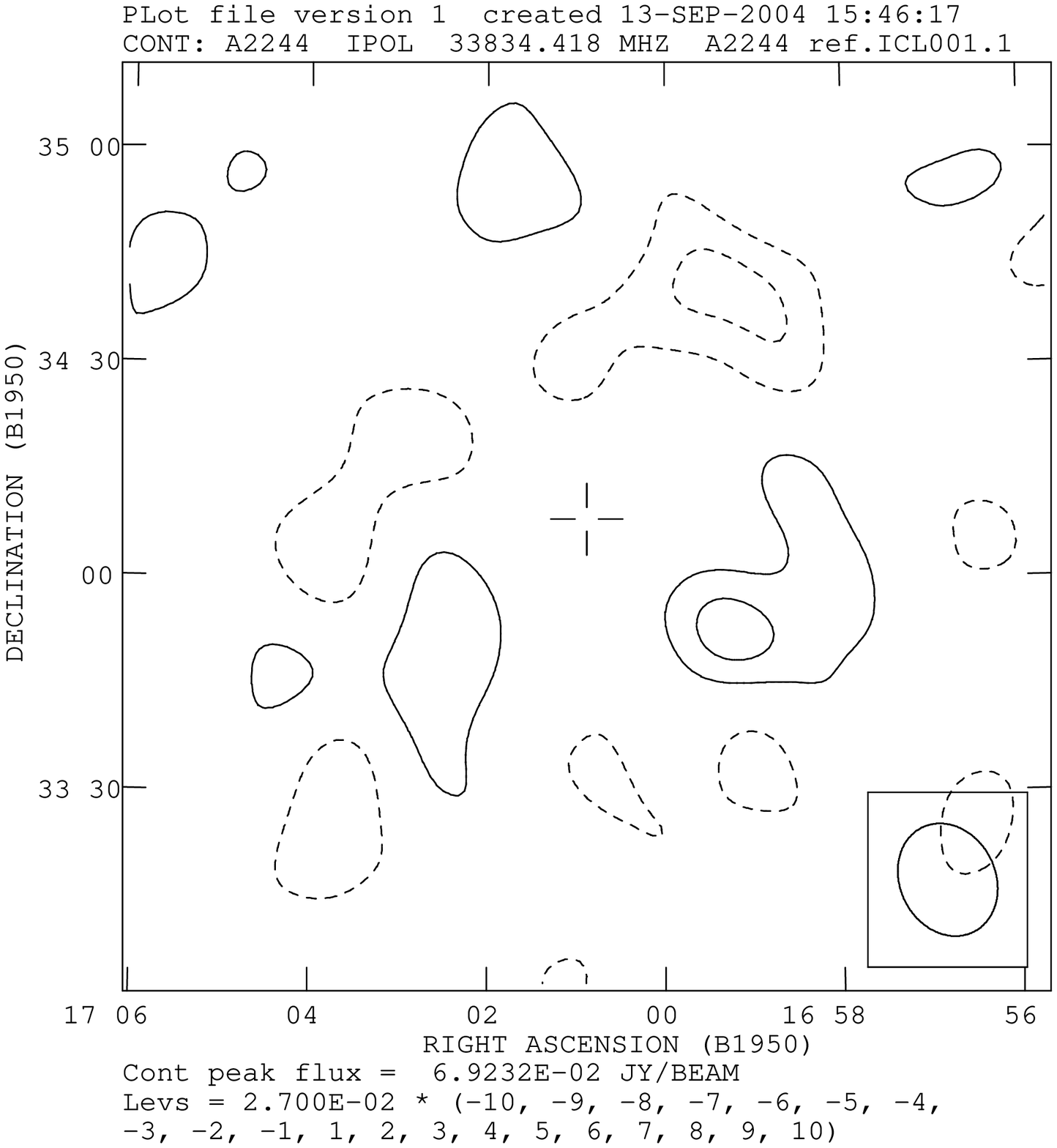}
  \caption{VSA maps of seven nearby clusters.  Top row, left to right:
    Coma, Abell~1795, Abell~399/401 (Abell~399 is further south).
    Bottom row, left to right Abell~478, Abell~2142 and Abell~2244.
    Coma, Abell~1795, Abell~478 and Abell~2142 are detected. 
    Contamination from the primordial MBR is evident in all
    maps. Contours are 1.5$\sigma$.} 
  \label{fig:vsa-maps}
\end{figure}

\subsection{Example interferometer results}\label{sec:intex}

The first interferometric map of the SZ effect, shown in
Fig.~\ref{fig:Ryle-SZ}, was made by Jones \etal\ (1993)
using a Ryle telescope observation of the cluster Abell~2218 which
lies at $z \approx 0.17$ (compare Fig.~\ref{fig:threeclusters}). When
making a map of such data, it is normal to include only the short
baselines where the SZ signal is strongest: longer baselines
contribute extra thermal noise, and have already been used to locate
and remove a number of confusing radio sources. The result of
censoring the baselines is that the maps have limited angular dynamic
range. The agreement of the SZ brightness recorded for Abell~2218 with
previous single-dish measurements established the credibility of SZ
effect research.

The BIMA and Owens Valley Radio Observatory (OVRO) arrays have also
made many SZ detections, including CL~0016+16 (BIMA) and a highly
significant detection of CL~0451-0354 (OVRO). Both are shown in
Fig.~\ref{fig:OVRO_BIMA-SZ} which was taken from Reese \etal\ (2000).
The high signal/noise SZ effect detections apparent in
Fig.~\ref{fig:OVRO_BIMA-SZ} can be achieved by long and repeated
integrations.

The obvious improvement that could improve the quality of SZ
observations is to design an interferometer to overcome the
difficulties of using ``normal'' interferometers. Baselines can be
tuned for optimum SZ effect detection over some redshift range
(i.e. range of angular sizes), and longer baselines can be added to
facilitate the removal of radio sources. An example of such a system
now in operation is the Very Small Array (VSA; Lancaster \etal\ 2004).

The VSA (Fig.~\ref{fig:vsa}), is a table-mounted interferometer run by
groups at the Cavendish Laboratory in Cambridge, Jodrell Bank
Observatory at the University of Manchester, and the Instituto de
Astrofisica de Canarias in Tenerife.  The telescope was specifically
designed to make observations of both primordial anisotropies in the
MBR and SZ effects from galaxy clusters.  Its range of baselines
($40\lambda < b < 300\lambda$ at 34\,GHz) make it particularly well
suited for observing low-redshift clusters, and observations of a
sample of low-redshift clusters are presented in Lancaster \etal\ (2004).
Fig.~\ref{fig:vsa-profiles} shows the predicted VSA
visibilities (flux density as a function of baseline; compare
Fig.~\ref{fig:profile}) for the seven $z < 0.1$ clusters in this X-ray
selected sample. 

Each cluster was observed for around 100 hours, resulting in some high
signal to noise detections, as shown for seven clusters in
Fig.~\ref{fig:vsa-maps}. All six maps appear noisy --- the noise
features are confusion from primordial MBR fluctuations, which are
relatively powerful on the range of angular scales defined by the
baselines used for these maps. Multi--wavelength observations capable
of removing the primordial MBR structures are needed for
improved sensitivity on these angular scales.

\begin{figure}
  \centering
  \includegraphics[width=0.9\linewidth,clip=]{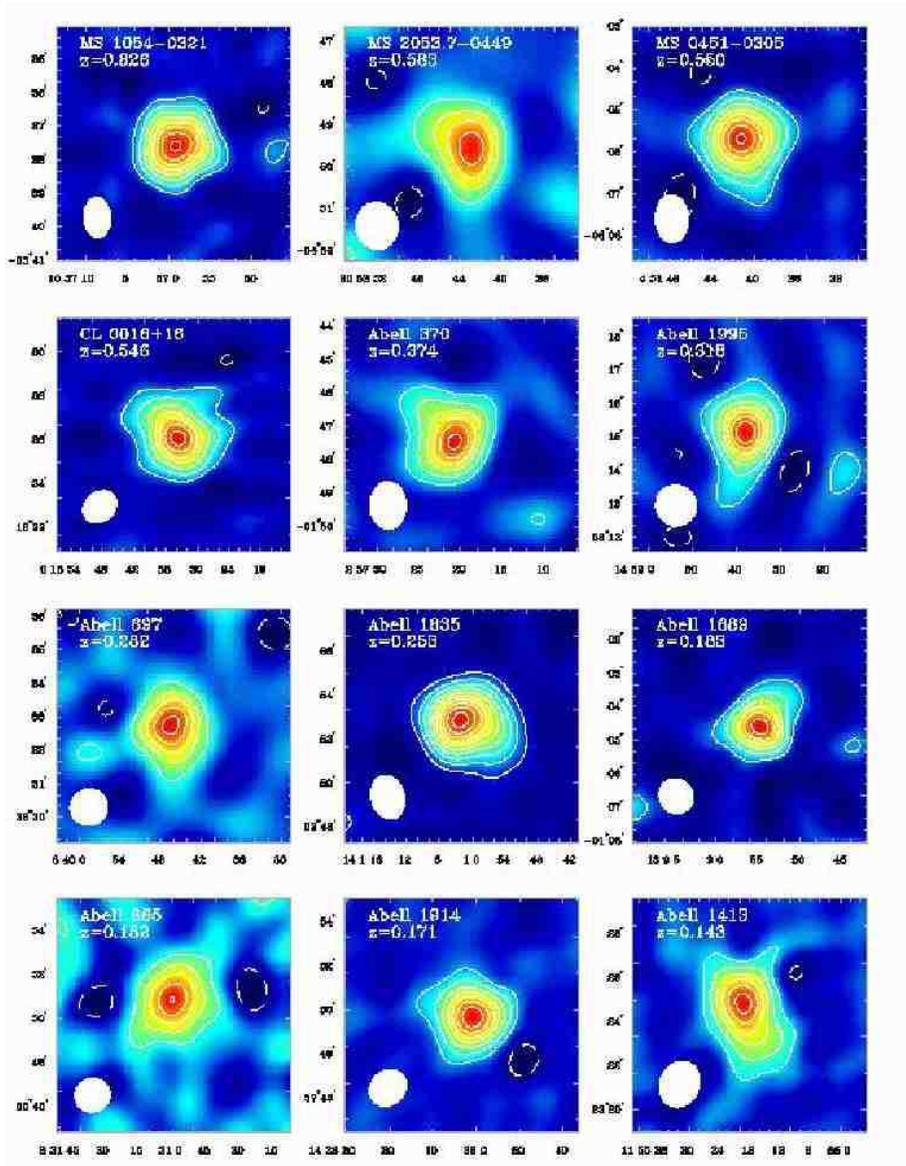}
  \caption{OVRO/BIMA maps of galaxy clusters in SZ.  Such observations
    are now becoming routine, often resulting in high--significance
    detections.} 
  \label{fig:many-sz}
\end{figure}

Interferometric SZ observations are now becoming almost routine, even
at $z \approx 1$. There have been around 50 cluster detections to
date, mostly with 
non-ideal telescopes, but these are more than adequate to the task.
Fig.~\ref{fig:many-sz}, taken from Carlstrom \etal\ (2002) shows
twelve high significance (better than $10\sigma$) examples --- many of
which have been confirmed by observations with multiple
telescopes. The wide range of redshifts of clusters in
Fig.~\ref{fig:many-sz} demonstrates the potential for using SZ effects
to study clusters throughout their histories.

\section{Systematic errors and the science return from
  Sunyaev-Zel'dovich effect studies} 

The SZ effects from clusters provide information on the properties of
those clusters, and can also be used to obtain important cosmological
information. The types of science that can be deduced from SZ effects,
and the sizes of the effects that need to be measured, are summarized
in Table~\ref{tab:szissues}. This section summarizes the procedures
needed to extract physics from SZ effect measurements, along with the
optical and X-ray data that are essential supplements, and how
uncertainties in the data impact the science that can be extracted.

\begin{table}
 \caption{Issues associated with SZ effect studies}
 \label{tab:szissues}
 \begin{narrowtabular}{16pt}{lcl}
  Use & Size of effect (mK) & Critical issues (Sections) \\
  \hline
  Cluster energetics                   & 0.5\phantom{1} 
          & \ref{sec:radcal} \\
  Cluster baryon count                 & 0.5\phantom{1} 
          & \ref{sec:radcal}, \ref{sec:isoth}, \ref{sec:model} \\
  Cluster gas structure                & 0.5\phantom{1} 
          & \ref{sec:beamshape}, \ref{sec:confusion} \\
  Cluster mass distribution            & 0.5\phantom{1} 
          & \ref{sec:radcal}, \ref{sec:isoth} \\
  Cluster radial peculiar velocity     & 0.05          
          & \ref{sec:radcal}, \ref{sec:confusion}, \ref{sec:bandpass},
            \ref{sec:velsubstruc} \\
  Cluster Hubble diagram               & 0.5\phantom{1} 
          & \ref{sec:radcal}, \ref{sec:model}, \ref{sec:clumping},
            \ref{sec:axial} \\
  Blind surveys and number counts      & 0.1\phantom{1}
          & \ref{sec:model}, \ref{sec:confusion} \\
  Baryon mass fraction evolution       & 0.1\phantom{1}
          & \ref{sec:radcal}, \ref{sec:isoth}, \ref{sec:model} \\
  Microwave background temperature     & 0.1\phantom{1}
          & \ref{sec:radcal}, \ref{sec:othersub} \\
  Cluster formation studies            & 0.02
          & \ref{sec:radcal} \\
  Cluster transverse peculiar velocity & 0.01
          & \ref{sec:confusion}, \ref{sec:polcal} \\
   \hline
 \end{narrowtabular}
\end{table}

\subsection{Cluster energetics}\label{sec:thermalen}

The total thermal SZ effect flux density, $S_{\rm th,\nu}$, from a cluster
usually dominates the kinematic effect, so that 
\begin{equation}
  S_{\nu} \approx S_{\rm th,\nu} \propto \int d\Omega \, \int
    n_e \, T_{\rm gas} \, dz \propto U_{\rm th}
  \label{eq:uth}
\end{equation}
where the integrals are over solid angle and distance along the line
of sight, and $U_{\rm th}$ is the total thermal energy content in the
atmosphere of the cluster. The constant of proportionality in
eq.~(\ref{eq:uth}) is composed purely of fundamental constants, the
thermodynamic temperature of the MBR, and
the frequency of observation, without any cosmological or structural
parameters, so that a measurement of the total SZ effect from a
cluster immediately provides a model-independent measure of the
thermal energy of the cluster gas.

If the gas is in hydrostatic equilibrium in the cluster gravitational
potential, then $U_{\rm th}$ should be closely related to the total
gravitational potential energy of the cluster, so that an SZ effect
survey should be able to pick out clusters of similar masses in
similar dynamical states at any redshift. With redshift information
from an optical follow-up programme, the evolution of cluster
potential wells could be studied and compared with the predictions of
cluster formation models. 

However, the usefulness of such SZ measurements is critically
dependent on the \textit{absolute} calibration of the SZ effect
data. 

\subsubsection{Absolute calibrations}\label{sec:radcal}

Calibrations of radio data (Secs~\ref{sec:singcal}, \ref{sec:intcal})
are often based on measurements of planets, and the conversion from
planet measurements 
at one frequency to the implied flux density from a planet at another
frequency depends on the properties of any planetary atmosphere, the
surface, and the polarization characteristics of the telescope as well
as the precision of the fundamental measurements of planetary flux
densities made using absolute instruments. 

Generally the complications in planetary
observations are relatively minor, so that the residual absolute flux
density errors depend on the original absolute calibrations of the
planet properties. The radio flux density scale used for SZ effect
observations is therefore good to about $5\%$, and this potential $5\%$
systematic error in the flux density scale should apply to all SZ
effect measurements to date.

Space-based observations of SZ effects, for example those to be made
by the \textit{Planck} satellite, should be better than this, since
they can be calibrated by reference to the MBR dipole anisotropy,
which is known from absolute measurements to higher accuracy. A
cross-calibration of the planetary scale against the MBR dipole would
then allow an improvement to all SZ effect measurements, and reduce
their systematic errors significantly.

\subsection{Cluster baryon count}\label{sec:baryons}

Eq.~(\ref{eq:uth}) implies that for a cluster with an isothermal
atmosphere
\begin{equation}
  S_{\rm th,\nu} \propto N_e T_{\rm gas}
  \label{eq:s-ne}
\end{equation}
where $N_{e}$ is the total number of electrons in the cluster. Thus
for an isothermal cluster, a model-independent estimate of the total
number of electrons in the cluster, and hence the total number of
baryons, can be deduced if the temperature, $T_{\rm gas}$,
and metallicity of the cluster gas can be deduced from a good X-ray
spectrum.

X-ray data for a cluster can also be used to calculate the cluster's
total mass, $M_{\rm tot}$, using the assumption of hydrostatic
equilibrium and spherical symmetry, eq.~(\ref{eq:equilmass}). If this
mass estimate is combined with the baryonic mass, taken from $N_e$
deduced from the SZ effect (or from the electron count deduced from
the X-ray data), the baryonic mass fraction in the cluster gas,
$f_{\rm b}$, can be found. Since the gas in clusters contains most of
the baryons (stars and stellar remnants are a relatively small
correction), the value of $f_b$ found in this way should be
appropriate for the cluster as a whole (and should 
be a reliable lower limit). If, then, clusters of galaxies are fair
samples of the mass content of the Universe, the derived value of 
$f_b$ should be close to the value $0.12 \pm 0.02$ (Turner 2002) for
the Universe as a whole.

On the other hand, if clusters are not fair samples of the total
matter content of the Universe, then the variation of $f_b$ as a
function of redshift, derived by this technique, would be a powerful
clue to the processes of cluster formation. At present there is no
evidence for significant variations in the cluster-based $f_b$ with
redshift (Carlstrom \etal\ 2002), but the errors on the values of
$f_b$ are large.

Clearly this measurement can be made only if the calibration of the SZ
effect is good (Sec.~\ref{sec:radcal}). The measurement also relies on
assumptions (Sec.~\ref{sec:isoth}) about the gas distribution.

\subsubsection{Isothermal spherical clusters}\label{sec:isoth}

The key assumptions about cluster properties in finding the baryon
count are
\begin{enumerate}
\item the gas must be isothermal --- this is hard to test, since the
 outer parts of the cluster are X-ray faint, but can make a
 significant contribution to the SZ effects; and
\item the measurement of the gas temperature must be good --- this
 depends on excellent calibration of the X-ray telescope, and also on
 the quality of the spectrum since the derived temperature is somewhat
 dependent on the metallicity of the cluster gas, and this is not
 usually well determined. For cooler clusters, the nature of the
 absorbing gas column in our Galaxy will also be important, since the
 shape of the X-ray spectrum at low energies depends on this column.
\end{enumerate}
The resulting systematic errors in the gas temperature (and it is
assumed that the electron and ion temperatures are the same: this is
by no means assured) are probably only $(1 - 2)\%$ in the X-ray bright
regions, so the appropriately-weighted systematic error in the gas
temperature over the population of electrons and baryons is likely to
be only $\sim 4\%$. This is not a major issue.

The strong additional assumption that the cluster has a spherical, or
almost spherical, shape is needed if the X-ray data are to be used to
determine the total cluster mass and so to determine the cluster
baryonic content. Projection effects that could affect the measurement
for $f_b$ are a major worry in using this method for individual
clusters: a preferable technique would be to apply it for clusters
selected in a manner independent of their orientation (i.e., using
clusters with surface brightnesses far above some X-ray threshold, or
using clusters from a blind SZ effect survey). The individual
measurements of $f_b$ would then be subject to projection effects, but
the average $f_b$ can be extracted from the population if the
intrinsic distribution of cluster atmosphere shapes can be recovered.

\subsubsection{Large-scale model}\label{sec:model}

The \textit{isothermal $\beta$ model} of Cavaliere \atque
Fusco-Femiano (1976) is a convenient and frequently-used description
of the large-scale structure of the atmosphere of a cluster of galaxies.
Its form can be derived from 
\begin{equation}
  G \, M_{\rm tot}(r) = - {k_B T_{\rm gas} r \over \mu m_{\rm p}}
    \, \left( {d\ln\rho_{\rm gas} \over d\ln r} + {d\ln T_{\rm gas}
    \over d\ln r} \right) \quad .
  \label{eq:equilmass}
\end{equation}
(e.g., Fabricant, Lecar \atque Gorenstein 1980)
which describes how the density, $\rho_{\rm gas}$, and temperature,
$T_{\rm gas}$, of a gas in hydrostatic equilibrium in a
spherically-symmetric gravitational potential well is related to the
mass of the cluster, $M_{\rm tot}(r)$, within radius $r$. 
If we assume that the gas is isothermal, and that the total mass
distribution has the form
\begin{equation}
  M_{\rm tot}(r) = 2 \, M_{\rm c} {r^3 \over r_{\rm c} \left(
                        r^2 + r_{\rm c}^2 \right)}
 \label{eq:mtot}
\end{equation}
where $r_{\rm c}$, the core radius, defines a characteristic scale
and $M_{\rm c}$ is the mass within $r_{\rm c}$, then a consistent
description of the density of the atmosphere is obtained with
\begin{equation}
  \rho_{\rm gas} = \rho_0 \left( 1 + {r^2 \over r_{\rm c}^2}
                   \right)^{-{3 \over 2}\beta} 
  \label{eq:beta}
\end{equation}
where $\beta$ is a constant which determines the shape of the gas
distribution, and depends on the ratio of a
characteristic gravitational potential energy and the thermal energy
in the gas 
\begin{equation}
  \beta = {2 \over 3} \, {\mu m_{\rm H} \over k T} 
                      \, {G M_{\rm c} \over r_{\rm c}} \quad .
\end{equation}
The alternative derivation of this isothermal $\beta$ model by
Cavaliere \atque Fusco-Femiano (1976) brings out the interpretation of
$\beta$ in terms of the relative scale heights of gas and dark matter
in the potential well. Most uses of eq.~(\ref{eq:beta}) seek to
determine $\rho_0$, $r_{\rm c}$, and $\beta$ without considering the
detailed properties of the underlying mass distribution.

It should be noted that the physical consistency of this much-used
model for the gas distribution depends on radial symmetry (a simple 
distortion from a spherical to an elliptical model for the gas density
$\rho$ would imply a mass distribution which is not necessarily positive
everywhere), and on gas at different heights in the atmosphere having
come to the same temperature without having necessarily followed the
same thermal history. Eq.~(\ref{eq:beta}) is not unique in
the sense that gas with a different thermal history might follow a
significantly different density distribution. Thus, for example, if
the gas has the same specific entropy at all heights, then the gas
density should be described by
\begin{equation}
  \rho_{\rm gas} = \rho_0 \, 
             \cases{
                \left( 1 - \alpha \ln \left
                     ( 1 + {r^2 \over r_{\rm c}^2} \right) \right)^{1
                     \over \gamma - 1}
                &
                $r \le r_c \left( e^{1/\alpha} - 1 \right)^{1 \over 2}$ \cr
                0
                &
                $r \ge r_c \left( e^{1/\alpha} - 1 \right)^{1 \over 2}$ \cr }
\end{equation}
where $\alpha$ is a structure constant with a similar meaning to
$\beta$. 

\begin{figure}
  \centering
  \includegraphics[width=0.5\linewidth]{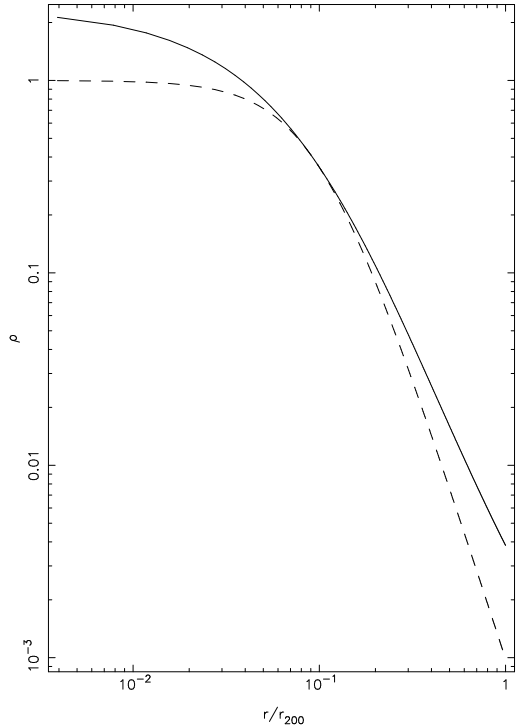}
  \caption{Representative isothermal models for cluster
    atmospheres. Solid line,  
    eq.~(\ref{eq:nfw}) with $r_s = 0.2 r_{200}$ and $\alpha =
    10$. Dashed line, eq~(\ref{eq:beta}) with $r_c = 0.1 r_{200}$ and
    $\beta = 1$.}
  \label{fig:nfwcompare}
\end{figure}

A similar procedure for the mass profile of Navarro, Frenk \atque
White~(1995), and an isothermal gas, leads to the gas density
distribution  
\begin{equation}
  \rho_{\rm gas} = \rho_{\rm s} \, 2^{-\alpha} \, \left( 1 + {r \over
         r_{\rm s}} \right)^{ \alpha { r_{\rm s} \over r}}
  \label{eq:nfw}
\end{equation}
where the new structure constant 
\begin{equation}
 \alpha = {1 \over \ln 2 - {1 \over 2}} \, {\mu m_{\rm H} \over k T}
   \, {G M_s \over r_s}  \quad .
\end{equation}
$r_s$ is the scale of the Navarro \etal\ model and $M_s$ is the
mass within radius $r_s$, so that $\alpha$ has a similar physical
meaning to $\beta$. It can be seen that in this solution $\rho
\rightarrow \infty$ as $r \rightarrow 0$. Examples of 
eq.~(\ref{eq:beta}) and~(\ref{eq:nfw}) profiles are shown in
Fig.~\ref{fig:nfwcompare}.

The total masses for both mass distributions in 
Fig.~\ref{fig:nfwcompare} diverge as $r \rightarrow \infty$, so
both must be truncated at some outer radius. One possibility is to
truncate at $r_{\rm 200}$, the radius at which the mean enclosed
mass density is $200 \times$ the critical density of the Universe
at the redshift at which the cluster is seen. 

The run of density and temperature in a cluster atmosphere are usually 
measured from the X-ray image and spectrum, where a density model for
the gas of the form of eq.~(\ref{eq:beta}) or eq.~(\ref{eq:nfw}) is
fitted to the X-ray surface brightness. The X-ray surface brightness
at a point offset by $r$ in {\it projected} distance
from the centre of the gas distribution (assumed spherical) is (in
energy per unit time per unit solid angle per unit frequency)
\begin{equation}
  \Sigma_{X}(r) = {\int \, n_{\rm e} n_{\rm p} \Lambda(T_{\rm gas}) dl
                      \over 4 \pi (1 + z)^3 } \quad .
\end{equation}
where $\Lambda(T_{\rm gas})$ is the X-ray emissivity of the gas. For
an isothermal gas with a $\beta$-model density distribution 
\begin{equation}
  \Sigma_{X}(r) = {\Lambda(T_{\rm gas}) \, n_{\rm e0} \, n_{\rm p0} \, r_c
                       \over 4 \, \pi \, (1 + z)^3 } \,
                       \left( 1 + {r^2 \over r_{\rm c}^2}
                       \right)^{{1 \over 2}-3 \beta} \,
                       \sqrt{\pi} \,
                       {\Gamma(3\beta - {1 \over 2}) \over
                       \Gamma(3\beta)} 
  \label{eq:sigmax}
\end{equation}
if the density distribution of eq.~(\ref{eq:beta}) is taken to extend to
infinity, with a constant gas temperature throughout.

This cannot be a fully physically-consistent description of the
atmosphere: it is clearly too simple, as would be the corresponding
result for the density distribution of eq.~(\ref{eq:nfw}). Both
involve infinite total masses, so an outer cut-off radius (not
consistently modelled) is adopted, and eq.~(\ref{eq:nfw}) also requires
an inner cutoff radius. The choice of an isothermal
description for the gas is also questionable: what process forces the
temperature to become constant, when the gas is accumulated at a
number of times from a number of sources (for example, infall and
galactic winds), and is imperfectly mixed by galaxy motions or
intracluster turbulence? 

However, some large-scale gas model like eq.~(\ref{eq:beta})
or~(\ref{eq:nfw}) is essential to relate the X-ray and SZ effects of
clusters since X-ray and SZ measurements are sensitive to 
different parts of the gas distribution. These models also provide a
convenient relationship between the atmosphere and the underlying mass
distribution. Note that eq.~(\ref{eq:beta}) and eq.~(\ref{eq:nfw})
generally require that the gas constitutes a radially-varying fraction
of the total cluster mass.

\subsection{Cluster gas structure}\label{sec:structure}

The structures of the atmospheres of clusters of galaxies are
generally studied by X-ray rather than SZ techniques at present, since
the signal/noise achievable in the X-ray is far superior. However, the 
X-ray emissivity is proportional to the square of the electron
density in the gas rather than to the electron density, so both the 
appearance of the X-ray image and the weighting of gas emissivity in
the X-ray 
spectrum are heavily influenced by the centre of the cluster. The
SZ effects, on the other hand, are directly proportional to electron
density, so they are relatively more sensitive to the outer parts of
the clusters and could be better probes of these regions. At
present the quality of the SZ data is such that only relatively crude
information on the structures of the atmospheres of nearby clusters is
available (e.g., Lancaster \etal\ 2004), but the redshift-independence
of the SZ effects could make them the best probes of the structures of
distant clusters. For the closest clusters, multi-wavelength studies
will be needed to remove the ``noise'' contributed by primordial
structure in the MBR.

For simple structural measurements using the thermal SZ effect no absolute
calibration is needed, and the X-ray data are useful only if the
conversion from the projected electron pressure profile to an electron
density profile is of interest. However the extraction of the
intrinsic profile from the observed profile relies on a deconvolution
with the telescope beam.

\subsubsection{Beamshape measurements}\label{sec:beamshape}

As discussed in Sec.~\ref{sec:singsel}, the telescope beamshape must be
well-known for accurate interpretation of the data. There are two
types of beamshape measurement that are needed to 
extract the best science from SZ effect data. First, since clusters
generally have a large angular size, they are likely to fill a
substantial fraction of the primary beam of an interferometer used to
observe the cluster (essentially, this is the same condition as the
requirement on the interferometer sampling for the observation to have
a high efficiency). This implies that the primary beams of the antennas
in the interferometer need to be well known if the SZ effect structure
is to be measured well on large angular scales.

On the other hand, small-scale structures are well represented in
interferometric data, and for these the beamshape is determined by the
sampling of the $u-v$ plane if the phase and amplitude errors in the
data are small. Thus small-scale SZ effect structures should be well
captured by interferometric observations and the deconvolutions
required to find their intrinsic properties should be reliable.

\subsubsection{Confusion}\label{sec:confusion}

Structural measurements of this type are clearly subject to confusion.
In Sec.~\ref{sec:intex} it was shown that primordial structures in
the MBR add a degree of noise to an SZ image that cannot be removed
without the use of multi-frequency observations
to effect a spectral separation of the thermal effect (and this brings
in, again, the requirement for high-quality spectral calibration;
Sec.~\ref{sec:radcal}). This imposes a practical limit on the angular
scales of SZ effect structure which can be studied at present.

At smaller angular scales, where the confusion from MBR structures is
reduced, confusion from non-thermal radio sources (principally quasars
and high-redshift radio galaxies) becomes important
(Fig.~\ref{fig:arcs}). Removal of most of the effect of these radio
sources is feasible using interferometers, as discussed in
Sec.~\ref{sec:intconf}, by the correct choice of
baselines or by aperture-plane fitting. However at the lowest SZ effect
flux densities it is hard to remove the confusing sources cleanly.

\subsection{Cluster mass distribution}\label{sec:lensing}

While it is possible to infer the total mass distribution of a cluster
using X-ray data and the assumption of hydrostatic equilibrium
(Sec.~\ref{sec:model}), a better approach may be to use weak
lensing. The advantage of this method is that a weak lensing
measurement of the ellipticity field induced by a cluster can be
converted into a map of the surface mass density of the cluster,
$\Sigma_{\rm tot}$, which is a linear projection of the total mass density
along the line of sight. The ratio of a thermal SZ effect map of the
cluster to this $\Sigma_{\rm tot}$ is then
proportional to the projected baryonic mass fraction along each line
of sight (if the cluster is isothermal). Simulations 
(Umetsu \etal\ 2004) show that blank-field SZ effect and lensing maps can
provide an estimate of the baryonic content and redshift of clusters
detected through either their shear field or their SZ effects.

This use of the SZ effect relies on the absolute calibration
(Sec.~\ref{sec:radcal}), in order to get the baryonic content, and the
assumption of isothermality (Sec.~\ref{sec:isoth}), just as the
SZ/X-ray approach to the baryon fraction. Otherwise this application
is subject principally to the problems of all lensing studies, for
example the question of the redshift distribution of the background
screen of galaxies and the possibility of multiple structures near the
line of sight. While these issues may affect the absolute value of the
baryonic content calculated, they should have little effect on the
image of the projected baryonic content that can be made, and so the
combination of SZ and lensing data may be a useful probe of the
large-scale gas-dynamical processes at work in clusters.

\subsection{Cluster radial peculiar velocity}\label{sec:velocity}

The use of the kinematic SZ effect to measure the radial velocity of a
cluster of galaxies relies heavily on the separation of the
kinematic and thermal SZ effects. This requires excellent relative
spectral calibration (Sec.~\ref{sec:radcal}) over a wide range of
frequencies --- LaRoque \etal\ (2004) fitted the spectrum of Abell~2163
over a decade of frequency in order to obtain their limit on its
radial peculiar velocity, and errors in the relative calibration cause
errors in the apparent spectral shape of the kinematic effect which
can have large effect on the radial velocity. 

Since the kinematic effect and primordial structure in the MBR have
the same spectrum, the kinematic effect is intrinsically confused by
the lumpiness of the MBR (Sec.~\ref{sec:confusion}). This confusion
limits the velocity accuracy that can be achieved on any single
cluster: the level of this limit depends on the angular scale being
examined, but exceeds about $200 \ \rm km \, s^{-1}$ for any plausible
present-day observation. Hence the statistical measurement of the
radial velocity distribution of clusters of galaxies, as derived from
a sample of clusters, is likely to be more useful than the measurement
of an apparent radial velocity of any single cluster.

\subsubsection{Bandpass calibration}\label{sec:bandpass}

Since the spectrum of the thermal effect is steeply rising through the
null at about 218~GHz, and observations of the intensity of the
combined thermal and kinematic effect in this part of the spectrum are
usually made using bolometers with wide spectral bandpasses, it is
clear that the bandpasses must be well known to avoid distorting the
apparent shape of the spectrum. Leakage of power into the detector
(principally from higher frequencies) could be a significant source of
problems.

\subsubsection{Cluster velocity substructure}\label{sec:velsubstruc}

While the kinematic SZ effect is usually discussed in terms of a
coherent motion of the entire cluster relative to the Hubble flow,
simulations generally show that the largest gas motions arise from
substructures within clusters. The largest of these substructures are
due to infalling groups, or the residual gas motions from
mergers. While this implies that there is significant small-scale
kinematic SZ effect structure, which is amenable to interferometric
study, the amplitude of individual effects is small, and the
superposition of the effects over a large fraction of the cluster
tends to produce a level of kinematic effect noise which again reduces
the accuracy of any measurement of the overall radial 
velocity of the cluster. Line-of-sight superpositions of these
substructures will also occur, and will tend to confuse their
details.

\subsubsection{Other substructure}\label{sec:othersub}

We should also note that the assumption of a smooth global mass
distribution leading to smooth density and temperature distributions
is certainly incorrect. There are a number of substructures found in
the X-ray images of clusters which affect the density and temperature
locally, and these will change the SZ effects \textit{to the extent
that the substructures produce pressure changes}, just as they alter
the X-ray appearances of clusters. The (thermal) SZ effect is
proportional to the integrated line-of-sight electron pressure, and so
a detailed SZ image of a cluster might show 

\begin{itemize}

\item pressure changes at cold fronts (e.g., as in Abell~2142;
 Markevitch \etal\ 2000); 

\item line-of-sight effects from cavities in the gas (e.g., near radio
 galaxies where radio-faint plasma bubbles are moving, or stationary,
 in the intracluster medium --- note that the non-thermal SZ effect
 from the relativistic electrons will be too small to compensate for
 the missing gas even if the bubbles are in pressure equilibrium;
 Birkinshaw 1999); 

\item pressure jumps from shocks around infalling groups or
 sub-clusters, as for example in the case of 1E~0657-56, where
 temperature and density jumps indicate a substructure moving at
 Mach~3 through the intracluster medium (Markevitch \etal\ 2002); 

\item variations in the static models eq.~(\ref{eq:beta},
 \ref{eq:nfw}) because of the slow changes and temperature gradients
 associated with cooling flows, which not only involve deviations from
 hydrostatic equilibrium but which are also likely to be non-steady; 

\item modest additional SZ effects from the non-thermal scatterings of
 populations of relativistic particles, particularly those associated
 with radio halo sources (Liang \etal\ 2002); and

\item kinematic SZ effects from motions of substructures in the
 clusters. 

\end{itemize}

\subsection{Cluster Hubble diagram}\label{sec:hubble}

One of the major uses to which the thermal SZ effect has been put is
that of determining the Hubble constant. In its simplest form, the
method is to compare the X-ray surface brightness and the thermal SZ
effect at the centre of a cluster 
\begin{eqnletter}
  \Sigma_{X,0}    &\propto& n_{e,0}^2 \, \Lambda(T_{\rm gas}) \, r_{c}
                            \\ 
  \Delta T_{RJ,0} &\propto& n_{e,0}   \, T_{\rm gas} \, r_{c}
\end{eqnletter}
where $r_{c}$ is some measure of the scale of the cluster (often the
core radius of the best-fitting isothermal $\beta$ model), and the
constants of proportionality include factors from the shapes of the
distributions of electron density and electron temperature. The X-ray
emissivity of the gas, $\Lambda(T_{\rm gas})$, is here taken to be a
constant over the atmosphere, assuming that the cluster is isothermal
and has a constant metal abundance. Then the combination
\begin{equation}
  \Delta T_{RJ,0}^2 \, \Sigma_{X,0}^{-1} \, \Lambda(T_{\rm gas}) \,
    T_{\rm gas}^{-2} \propto r_{c}
  \label{eq:rcore}
\end{equation}
and a measure of the angular size of the cluster, $\theta_{c}$,
can be compared with the linear size, $r_{c}$, to derive the angular
diameter distance of the cluster
\begin{equation}
  D_{A} \propto \Delta T_{RJ,0}^2 \, \Sigma_{X,0}^{-1} \,
    \Lambda(T_{\rm gas}) \,
    T_{\rm gas}^{-2} \, \theta_{c}^{-1} \quad .
  \label{eq:dist}
\end{equation}

An example of the application of this method is given in
Sec.~\ref{sec:0016}, but it is clear from eq.~(\ref{eq:rcore}) that
errors in the SZ effect and X-ray temperature have a major impact
on the accuracy of the angular diameter distance that is derived.
Although the angular diameter distances derived in this way can be
used to construct a ``cluster Hubble diagram'' of angular diameter
distance as a function of redshift (Birkinshaw 1999; Carlstrom 
\etal\ 2002), and in principle this diagram could be used to deduce
further 
cosmological parameters than the Hubble constant, the errors on the
distance estimates are currently too large to make this possible.

Clearly a major requirement for the use of this method is that the
thermal SZ effect and the X-ray brightness of the cluster are accurately
determined: the dominant error is from the SZ effect, since a 5\%
systematic error in the flux density scale of the SZ effect
(Sec.~\ref{sec:radcal}) leads to a 10\% error in the distance scale. 
However, if the Hubble constant itself is not of interest, but rather
the intention is to use the Hubble diagram to estimate the density
parameters $(\Omega_{m 0},\Omega_{\Lambda 0})$, then the absolute
calibration not important --- it represents merely a shift of the
overall distance scale --- and only the shape of the angular diameter
distance/redshift relation is important. 

The large-scale model of the cluster gas (Sec.~\ref{sec:model})
determines the constant of proportionality in eq.~(\ref{eq:dist}), 
and so it is essential that a correct model is adopted. Since the
X-ray data are sensitive to emission from the inner part of the
cluster, while the SZ data are relatively more sensitive in the outer
regions (because of the different $n_e$ scalings), a
high-sensitivity X-ray image of the cluster is essential to trace the
gas out to sufficient distance that the model can be used with
confidence.

\subsubsection{Clumping}\label{sec:clumping}

On the smallest angular scales, below the angular resolution of the
X-ray image, the gas may be significantly clumped because of the
effects of turbulence induced by galaxies moving at transonic speeds,
the dissipation of gas from infalling groups, local heating from
low-power radio sources, etc. Such clumping has a significant
effect on the constant of proportionality in eq.~(\ref{eq:dist}). For
example, if the clumping is isothermal, then the clumping factor
\begin{equation}
  C = {\langle n_e^2 \rangle \over \langle n_e \rangle^2}
\end{equation}
measures the excess of X-ray emissivity over that obtained from a
smooth density distribution, and the distance inferred by assuming
that the density distribution is smoothed and unclumped is an
underestimate by a factor $C$. Limits to the amount of non-isothermal
clumping can be deduced from the detailed X-ray spectrum of a cluster,
if the spectrum contains enough counts, but limits on the amount of
isothermal clumping can only be based on theoretical considerations
about the dissipation time of the implied overpressure and the rate of
creation of the clumps.

This issue may induce not only a scale error in the Hubble diagram,
and hence an error in the Hubble constant, but could also change the
shape of the diagram if the average dynamical state of the
intracluster medium evolves with time --- perhaps from a clumpy
initial state, just after the atmosphere assembles, into a smoother
and more relaxed state at the present time. Thus the uncritical use of
the cluster Hubble diagram may lead to significant errors in the
estimation of the cosmological parameters $(\Omega_{\rm m 0},
\Omega_{\Lambda 0})$ that dictate how the angular diameter distance
changes with redshift.

\subsubsection{Axial ratio}\label{sec:axial}

Eq.~(\ref{eq:dist}) derives the angular diameter distance for a
cluster by comparing its line-of-sight scale with its transverse
angular size. If the cluster is non-spherical, then this ratio will
not yield the angular diameter distance correctly. However, if a
sample of clusters at similar redshifts, with random orientations, is
used, then an average over this sample with its various cluster shapes
should reduce the error, at the expense of adding substantially to the
noise in the distance estimate. 

Such an average will not be successful if the set of clusters is
biased, and a bias is likely for the faintest clusters since
non-spherical atmospheres have higher central surface brightnesses,
and so are easier to detect, if their long axes lie close to the line
of sight. It is, therefore, important to select a sample of clusters
that has no orientation bias. This can be done either by selecting
clusters which are far above the surface brightness limit of some
finding survey, or by selecting clusters based on an integrated,
surface-brightness independent, indicator of cluster properties. An
ideal selection would be the integrated thermal SZ effect, since this
is a linear indicator of the total electron count in a cluster, and so
is orientation independent.

\subsection{Blind surveys and number counts}\label{sec:surveys}

SZ-effect selected samples of clusters of galaxies would be ideal for
many cosmological purposes, not only for the measurement of the
distance scale. This is because the SZ effects are redshift
independent in surface brightness terms, so that clusters can be
detected to high redshift. An SZ effect survey could therefore be
constructed to be 
almost mass limited (as shown by the flat efficiency curve in
Fig.~\ref{fig:zdep}), as well as being orientation independent. 

A sample of clusters found by such a survey should therefore provide a
fairly direct indicator of 
how many clusters of a given mass have assembled at any redshift, and
the cluster count and redshift distribution can be used to set strong
constraints on $\sigma_8$ and $\Omega_{\rm m 0}$ (e.g., Fan \atque Chiueh
2001) provided that our understanding of the early phases of cluster
formation is accurate. Alternatively, we could regard the cluster
statistics as a test of our models of cluster formation.

Large surveys for clusters in their SZ effects are only now beginning
to be made: up until the present, the sensitivity of interferometers,
radiometer arrays, and bolometer arrays has been insufficient to allow
surveys of a sufficiently large area of sky for useful statistics on
the cluster population to be developed. Perhaps the fastest approaches
involve the use of radiometer and bolometer arrays, and the best areas
to survey are those which already have significant optical and X-ray
coverage (since confirmation of potential cluster detections is a
crucial aspect of the work).

Clearly, the absolute calibration of the SZ effects is not a critical
issue if the aim of blind surveys is merely to detect clusters, but
the surveys will have to contend with the problems of radio source and
MBR confusion (Sec.~\ref{sec:confusion}). Interpreting the results of
the survey must rely on a good knowledge of the selection function:
the efficiency of cluster detection as a function of redshift in the
presence of MBR fluctuations, radio sources (with a population which
evolves with redshift), etc.. The calculation of this selection
function will depend on the accuracy of the cluster model
(Sec.~\ref{sec:model}).

\subsection{Baryon mass fraction evolution}\label{sec:baryonevol}

This application of SZ effects has been dealt with earlier, for
individual clusters (Sec.~\ref{sec:baryons}). The additional difficulty
for examining the evolution is only that of knowing the cosmological
parameters sufficiently well that the investigation of clusters at
different redshifts can work to a constant metric radius, or
alternatively to some other well-defined radius (such as $r_{200}$).

\subsection{Microwave background temperature}\label{sec:mwbt}

The ratio of the thermal SZ effect of a cluster at two different
frequencies is a function of the temperature of the MBR, with some
slight dependence on the temperature of gas in the cluster and the
cluster radial velocity. Thus a precise measurement of the SZ effect
spectrum can be used to measure the MBR temperature at distant
locations, and over a wide range of redshifts. This would allow a test
of the MBR temperature evolution
\begin{equation}
  T_{\rm rad}(z) = T_{\rm rad}(0) \, (1 + z)
  \label{eq:tradz}
\end{equation}
expected from our normal model of radiation in the Universe. 
Battistelli \etal\ (2002) found no deviation from eq.~(\ref{eq:tradz}),
but the errors on the SZ effect amplitudes they used were substantial,
as were the errors on the MBR temperatures at high redshift (which
were obtained from molecular excitation studies). At present this
method is relatively insensitive to deviations from
eq.~(\ref{eq:tradz}). 

Clearly this technique relies on accurate knowledge of the spectrum of
the thermal SZ effect, and therefore on the absence of any significant
kinematic SZ effect. Filtering of the effects requires precise
knowledge of the relative calibration of the SZ effect data at widely
separated frequencies (Sec.~\ref{sec:radcal}), and the absence of
temperature substructures in the cluster (Sec.~\ref{sec:othersub})
large enough to cause significant spectral changes.

\subsection{Cluster formation studies}\label{sec:form}

The SZ effects related to the formation of clusters should be those of
lowest brightness and smallest angular size, since they are generated
as cluster potential wells begin to accumulate gas, and that gas is
heated by the impacts of infalling groups. The angular scales that are
relevant are of order $10$~arcsec, and the amplitudes of the effects
are $\sim 20 \ \rm \mu K$. Such structures are only detectable
against the background of MBR confusion because they have a rising
power spectrum while primordial MBR fluctuations have a falling
spectrum at such small scales (Molnar \atque Birkinshaw 2000), and
because it should be possible to use spectral techniques to separate
the thermal SZ effect from primordial fluctuations.

Efforts to look at the first clusters are generally better in the SZ
effects than in the X-ray: the SZ effects begin to increase in flux
density beyond $z \sim 1.6$ because of the redshift-independence of
their surface brightnesses and the behaviour of $D_{\rm A}(z)$, while
it is already difficult to detect high-mass X-ray clusters beyond $z =
1$. 

The prospects of detecting the SZ effects of the first clusters are
not currently good, since even allowing for the efficiency with which
SZ effect surveys can find
distant clusters, the amplitude of the thermal signal is low. Indeed,
for sufficiently fast infall and sufficiently cool accreting groups,
the principal detectable SZ effect may be kinematic rather
than thermal, and then the separation from primordial
fluctuations in the MBR can only be done statistically (for example
using deviations from the expected primordial power spectrum).

Crucial requirements for attempting this work are, therefore, the
ability to attempt accurate spectral separation (and hence the
requirement for good absolute calibration; Sec.~\ref{sec:radcal}), 
and the development of a suitable telescope for this work
(Sec.~\ref{sec:iszo}).

\subsection{Transverse velocities}\label{sec:vtrans}

The transverse velocities of clusters can be extracted from the
comparison of the amplitude and polarization of their thermal SZ
effects. Even for the brightest SZ cluster, the amplitude of the
polarization signal is expected to be under $10 \ \rm \mu K$, and this
polarization (with a uniform direction across the cluster) must be
separated from the circumferential pattern of polarization induced by
multiple scatterings (Sec.~\ref{sec:radquan}). 

Since many of the radio sources contributing to the confusing
background are themselves polarized, a first requirement for this
observation to be made is that the confusion be well controlled
(Sec.~\ref{sec:confusion}).

\subsubsection{Polarization calibration}\label{sec:polcal}

It is clear that excellent polarization calibration of the
equipment used will be essential. Attempting to detect sub-$\mu \rm
Jy$ polarization signals in the presence of mJy~signals from the
thermal SZ effect and similar or larger signals from confusing radio
sources 
will require the polarization behaviour of the telescope to be well
controlled and the cross-polarization calibrations to be unusually 
precise. No telescope currently existing is capable of making this
measurement on the interesting angular scales: an exploration of the
possibilities of polarization measurements may be possible with the
next generation of SZ effect telescopes (Sec.~\ref{sec:iszo}).

\section{A case study: CL~0016+16}\label{sec:0016}

In this section, we discuss in more detail the procedures that are
undertaken to combine SZ effect measurements with other data to
extract cosmological and astrophysical information, by reference to
the cluster CL~0016+16.

CL~0016+16 was discovered as a candidate high-redshift ($z \sim 0.5$)
cluster by Kron (1980). Its basic characteristics in the optical,
X-ray, and SZ effect were described by Koo \etal\ (1981), White 
\etal\ (1981), and Birkinshaw \etal\ (1981) in companion papers. Although
the initial discovery was in the optical, this can also be regarded as
an X-ray selected cluster, since it appeared independently in the
\textit{Einstein} Medium-Sensitivity Survey (Gioia \etal\ 1990).

\begin{figure}
  \centering
  \includegraphics[width=0.8\linewidth]{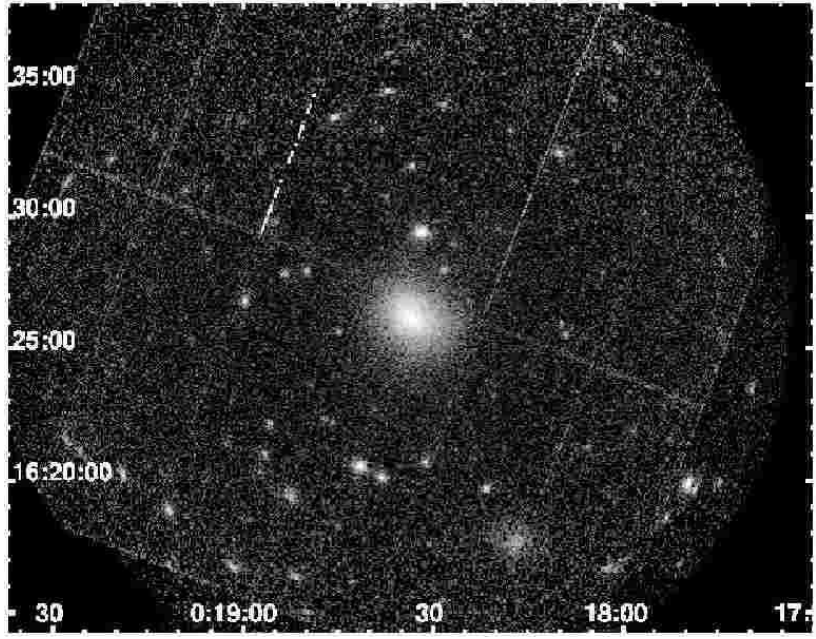}
  \caption{Combined MOS-1, MOS-2, and pn image of CL~0016+16 in 0.3~to
    5.0~keV, corrected for vignetting but without background
    subtraction. Artefacts associated with the edges of the chips are 
    evident. CL~0016+16 is the dominant central extended
    source. The associated quasar and one of the associated companion
    clusters lie just north and to the south-west, respectively, of
    CL~0016+16.}
  \label{fig:cl0016xmmimage} 
\end{figure}

Since CL~0016+16 has high optical and X-ray luminosity, and a
substantial SZ effect, it has been the subject of a considerable
number of studies since the early 1980s. Recent measurements show it
to have a redshift of $0.5481$ (Ellingson \etal\ 1988), and to be the
dominant cluster in a supercluster with at least two adjacent X-ray
clusters of lower mass (Hughes \etal\ 1995, Hughes \atque Birkinshaw
1998). A quasar of similar redshift (Margon \etal\ 1983) also seems to
be associated with this supercluster. CL~0016+16 has a
$0.5 - 4.5$~keV X-ray luminosity of about $2 \times 10^{38} \ \rm W$.

\subsection{X-ray data}

\textit{ASCA} and \textit{ROSAT} studies of CL~0016+16 showed the
cluster to have a somewhat elliptical shape and a gas temperature of
about 7.6~keV (Hughes \atque Birkinshaw 1998). However the errors on the
structural parameters of the gas ($\beta$ and $\theta_{\rm c}$) in
(\ref{eq:beta}) and on the temperature remained a limiting factor in
the interpretation of the cluster. \textit{XMM-Newton} observations of
the cluster were therefore made to provide better measurements of the
properties of the atmosphere. A full description of the treatment of
the data is given in Worrall \atque Birkinshaw (2003).

\begin{figure}
  \centering
  \includegraphics[width=0.5\linewidth]{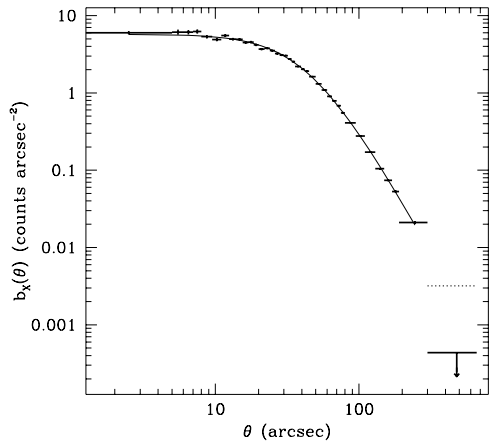}
  \caption{The 0.3~to 5.0~keV radial profile of CL~0016+16 after
    subtraction of local background. The curve shows the best-fit
    circularly symmetric isothermal $\beta$ model convolved with the
    point spread function of XMM. The horizontal dotted line shows the
    range of radii used for local background, and the level of that
    background.}
  \label{fig:cl0016radial} 
\end{figure}

The \textit{XMM-Newton} observation of CL~0016+16 took 37~ks of data
of which rather little was lost to particle flares. An image of the
cluster formed from the combined dataset from the three cameras on XMM
is shown in Fig.~\ref{fig:cl0016xmmimage}.
The overall structure of the cluster is somewhat elliptical, but
relatively smooth. To first order in the ellipticity we can take the
cluster to be circularly-symmetric, and fit the radial profile using
the isothermal $\beta$ model, eq.~(\ref{eq:beta}). The fit is good
(Fig.~\ref{fig:cl0016radial}), and yields structural parameters
$\beta = 0.70 \pm 0.01$ and a core radius $\theta_{\rm c} = 36 \pm
1$~arcsec (corresponding to a linear core radius $r_{\rm c} = 240 \pm
10 \ \rm kpc$ in the standard cosmological model).

\begin{figure}
  \centering
  \includegraphics[width=0.5\linewidth]{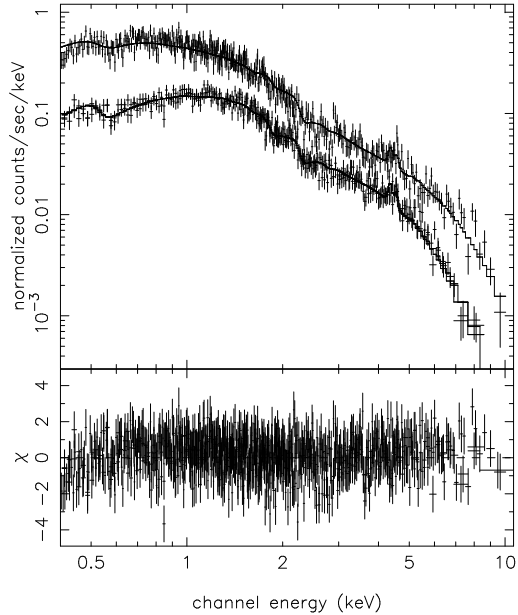}
  \caption{The XMM spectrum of CL~0016+16 from a circle of radius
    90~arcsec. The upper spectrum is derived from the pn data, while the
    lower is from the MOS-1 and MOS-2 data. The fit is to an isothermal
    gas with $k_{\rm B} T = 9.1$~keV, and abundance of 0.22~times the
    solar value, at the optical redshift of the cluster.}
  \label{fig:cl0016xmmspec} 
\end{figure}

A spectrum of CL~0016+16 extracted from the central part of the
cluster is shown in Fig.~\ref{fig:cl0016xmmspec}. This spectrum can be
well fitted by a single-temperature plasma, with $k_{\rm B} T_{\rm gas}
= 9.1 \pm 0.2$~keV and an abundance of $0.22 \pm 0.04$~times the solar
abundance. There are sufficient counts in this spectrum (32600~net
counts from the three cameras in $0.3 - 10.0$~keV) that the
redshift of the emitting plasma can be determined, and shown to be
consistent with the optically-derived redshift of the cluster. It
should be noted that there remain uncertainties in the spectral
extraction and fitting procedure because of the significant
backgrounds in the X-ray images and the complicated distributions of
background counts in energy and position on the cameras, and because
of residual uncertainties in the calibration.

A further issue is that of structural or thermal substructure in the
cluster. A close examination of the central part of the cluster's
X-ray image reveals a small central sub-structure that is also seen in
the \textit{Chandra} data. The change in central brightness in the radial
profile Fig.~\ref{fig:cl0016radial} due to this central structure is
small, so the overall parameters of the $\beta$ model shouldn't be
affected by its presence, but there is some evidence that the central
structure is slightly ($\sim 0.5$~keV) hotter than the bulk of the
cluster, although the significance is not high.

There are some other variations from the $\beta$ model profile: in
particular there is a region of low surface-brightness emission to the
SW of the cluster that is also seen in an earlier ROSAT image (Neumann
\atque B\"ohringer 1997), but again these brightness structures are
too small to affect the overall structural fits.

With the aid of the gas temperature and abundance that are provided by
the analysis of the X-ray spectrum, the detected counts can be
converted into the central density of the gas in the cluster
(conveniently characterised by the central proton density, $n_{\rm
p0}$) if some assumption is made about the cosmology. The derived
$n_{\rm p0}$ and $k_{\rm B} T_{\rm gas}$ are subject to systematic
errors from background subtraction uncertainties in both the radial
profile and the spectrum, and from our ignorance about small-scale
temperature variations in the cluster. 

\subsection{SZ data}

The SZ effect of CL~0016+16 has been reported many times using a
number of different techniques: the large amplitude and angular size
of the effect make it a useful test case for new SZ observational
programmes. Three recent results are reported in
Table~\ref{tab:cl0016sz}. Of these measurements, one used a large
single-dish system equipped with a dual-beam radiometer (Hughes \atque
Birkinshaw~1998), and the other two used interferometers designed for
other purposes which had been retro-fitted to provide high sensitivity
on large angular scales (Reese \etal\ 2000; Grainge \etal\ 2001). 

\begin{table}
 \caption{SZ effect data for CL~0016}
 \label{tab:cl0016sz}
 \begin{narrowtabular}{8pt}{llc}
  Paper & Telescope & $\Delta T_{RJ}$ (mK) \\
  \hline
  Hughes \atque Birkinshaw (1998) & OVRO 40-m telescope
                                  & $-1.20 \pm 0.19$ \\
  Reese \etal\ (2000)             & BIMA and OVRO interferometers
                                  & $-1.24 \pm 0.11$ \\
  Grainge \etal\ (2002)           & Ryle telescope 
                                  & $-1.08 \pm 0.11$ \\
  \hline
  \multispan{2}{\hfil Converted to XMM model} & $-1.26 \pm 0.07$ \\
  \hline
 \end{narrowtabular}
 
\end{table}

Although CL~0016+16 has one of the largest SZ effects, in terms of the
central surface brightness, the measurements have relatively low
signal/noise compared with the X-ray data. $\beta$ model fits to the
SZ data are consistent with the fits obtained from the X-ray data, but
add little weight towards improving the errors on the derived structural
parameters. Thus it is better to use the X-ray derived structural
parameters to fit the SZ data (or to undertake a simultaneous SZ and
X-ray fit) than to perform independent fits. 

Since the three SZ effects reported in Table~\ref{tab:cl0016sz} were
extracted from the raw data via the application of a different model
for the gas, and each model is somewhat different from the model
fitted from the X-ray data, it is necessary to correct them to the
same overall measure of the SZ effect. The overall result after
scalings that take account of the different efficiencies of the
observing techniques is that the central SZ effect, at zero frequency,
is $-1.26 \pm 0.07$~mK. 

\subsection{Distance of CL~0016+16}\label{sec:cldist}

We can apply the formalism derived in the discussion of the Hubble
diagram (Sec.~\ref{sec:hubble}) to estimate the distance of the
cluster. The X-ray emission measure of CL~0016+16, integrated over the
entire best-fitting $\beta$ 
model, is
\begin{equation}
  { \int n_{\rm e} n_{\rm p} dV \over 4 \pi D_{\rm L}^2}
  =
  (1.75 \pm 0.05) \times 10^{21} \quad \rm m^{-5}
\end{equation}
and for the isothermal $\beta$ model we expect that
\begin{equation}
  { \int n_{\rm e} n_{\rm p} dV \over 4 \pi D_{\rm L}^2}
  =
  {n_{\rm e0}^2 D_{\rm A}} \, {\theta_{\rm c}^3 \over \eta \, ( 1 +
  z)^4} \,
  {\sqrt{\pi} \over 4} \, {\Gamma\left( 3\beta - {1 \over 2} \right)
  \over \Gamma\left( 3\beta \right)}
  \label{eq:em2}
\end{equation}
(as in eq.~(\ref{eq:sigmax}), 
where $\eta$ is the electron/proton ratio, taken as being constant
over the entire cluster. Since the X-ray data measure $\beta$ and
$\theta_{\rm c}$ from the structure of the atmosphere, $\eta$ (which
depends on the abundance of metals in the gas) and $T_{\rm gas}$ from
the X-ray spectrum, and the redshift is available either from the
X-ray spectrum or optical measurements, the emission measure implies a
value for the combination $n_{\rm e0}^2 D_{\rm A}$ of the unknown
central gas density in the cluster and its angular diameter
distance.

The central SZ effect of $\Delta T_{RJ,0} = -1.26 \pm 0.07$~mK measures a
different combination of gas properties and distance,
\begin{equation}
  \Delta T_{RJ,0}
  =
  - {n_{\rm e0} D_{\rm A}} \, 
  {2 \sigma_{\rm T} \, T_{\rm rad} \, \theta{\rm cx}} \,
  {k_{\rm B} T_{\rm gas} \over m_{\rm e} c^2} \,
  \sqrt{\pi} \, {\Gamma\left( {3\over 2} \beta - {1 \over 2} \right)
  \over \Gamma\left( {3\over 2}\beta \right)}
\end{equation}
and so $n_{\rm e0} D_{\rm A}$ can be derived from the SZ data. These
results can therefore be combined to produce measurements of $D_{\rm
A}$ and $n_{\rm e0}$. 

The distance of CL~0016+16 that results is
\begin{equation}
  D_{\rm A} = 1.36 \pm 0.15 \quad \rm Gpc
  \label{eq:darandom}
\end{equation}
where the error, at present, is random only. The largest contribution
to this random error come from the error on the amplitude of the SZ
effect (about 6\%, which leads to a 12\% error in the distance): an
improved measurement of the SZ effect in even this cluster, which has
a relatively well-measured SZ effect, would immediately lead to a
significant improvement in distance. However, the major issue in
absolute distance measurements of this type is not the random error,
which could be reduced by continued observations or similar-depth
observations of a large number of clusters, but rather the systematic
error. 

Similarly, a cosmology-independent central electron density can be
deduced by combining the X-ray and SZ effect data. The result
\begin{equation}
  n_{\rm e0} = (8.8 \pm 0.5) \times 10^3 \quad \rm m^{-3}
\end{equation}
has a significantly smaller random error than the distance because of
the lower power of the SZ effect normalization that appears in the
expression for $n_{\rm e0}$. 

The systematic errors in the angular diameter distance that arise in
the application of this method are summarized in
Table~\ref{tab:distsyst}. The major issue for any individual cluster
is that of its unknown 3-D shape. Even if we take cluster shapes as
being simple oblate or prolate ellipsoids, the possible variations in
the estimated distance can be large (e.g., Hughes \atque Birkinshaw
1998). Additional uncertainty in the X-ray and SZ effect form factors
arises because the X-ray measurements are principally sensitive to 
the central cluster emission (since X-ray emissivity depends on
$n_{\rm e}^2$), while the SZ effect is proportional to
$n_{\rm e}$, and so is relatively more sensitive to the outer
regions. Thus changes in shape between the inner and outer parts of
the cluster can produce an error in the distance which is effectively
unprobed by the X-ray profile even if the structure is
spherically-symmetric. For CL~0016+16, where an unusually deep
X-ray observation exists, the X-ray data probe to angles $< 5$~arcmin
from the centre, so that about $8$~per cent 
of the central SZ effect
originates from larger radii and so is represented in the model used
to estimate distance only through an extrapolation of the $\beta$
model. The fraction of the gas that is unprobed is usually
substantially larger than this, leading to a larger potential problem
in relating the X-ray and SZ effect data to determine the distance.

Small-scale thermal and density substructure within the cluster can
also affect the estimate of distance by boosting the X-ray output
relative to the SZ effect (for example by isobaric clumping of the
intracluster medium), or by boosting the SZ effect relative to the
X-ray output (for example through the existence of regions of
shock heating which can cause only small changes in the X-ray spectrum
but large local enhancements in electron pressure). The upper limit on
the amplitude of these effects in Table~\ref{tab:distsyst} is believed
to be conservative, and could be reduced by improved X-ray
spectroscopy. However, in a sense this problem could be improved by
observing a sample of clusters, since it is unlikely that each will
have the same level of substructure. Major outliers 
could then be detected and removed. For any individual cluster, only
simple checks against the presence of substructure in the image (via
deviations from Poisson statistics) or in the spectrum (via
introducing a distribution of emission measures corresponding to
different temperatures) can be made. 

\begin{table}
 \caption{Systematic errors in distance measurements}
 \label{tab:distsyst}
 \begin{narrowtabular}{32pt}{llc}
  Class of error & Component & $\sigma_{\rm D}/D_{\rm A}$ \\
  \hline
  Calibration   & SZ effect amplitude         & $\phantom{<} 0.10$ \\
                & X-ray emission measure      & $\phantom{<} 0.05$ \\
                & X-ray temperature           & $         <  0.02$ \\
  Projection    & 3-D shape of cluster        & $\phantom{<} 0.20$ \\
  Substructure  & Clumping, shocks, etc.      & $\phantom{<} 0.15$ \\
  Contamination & Kinematic SZ, sources, etc. & $\phantom{<} 0.05$ \\
  \hline
  \multispan{2}{\hfil Quadrature combination} & $\phantom{<} 0.25$ \\
  \hline
 \end{narrowtabular}
\end{table}

Absolute distance measurements like this require absolute calibration
of the data used to make the measurements, and so the ultimate
reliability of the distances will depend on the reliability of the
calibrations. The largest problem is likely to be in the calibration
of the radio flux density scale (as described in Sec.~\ref{sec:radcal}),
and efforts to improve the calibration of the scale at each of the
frequencies used for SZ effect measurement are essential if the
distances are to be free from systematic shifts in the distance scale
which cannot be identified by sample studies. 

If we combine the random and systematic errors on the distance, then
the updated result for the angular diameter distance of CL~0016+16
from eq.~(\ref{eq:darandom}) is
\begin{equation}
  D_{\rm A} = \left ( 1.36 \pm 0.15 \right) \pm 0.34 \quad \rm Gpc
  \label{eq:dafinal}
\end{equation}
where the first error is random, and the second systematic. This can
be converted into a 
measurement of the Hubble constant if we make some assumption about
the correct cosmology. Adopting a flat cold dark matter cosmology with
a cosmological constant contributing 70\% of the closure density, we
find 
\begin{equation}
  H_0 = \left( 68 \pm 8 \right) \pm 18 \quad \rm km \, s^{-1} \, Mpc^{-1}
\end{equation}
which is consistent with the consensus distance scale. The larger
value than found in earlier results for CL~0016+16 (which gave a
result near $50 \ \rm km \, s^{-1} \, Mpc^{-1}$) arises almost equally
from an increase in the temperature of the gas (from 7.5 to 9.1~keV)
and from the change from an open CDM to a flat $\Lambda$CDM cosmology.

\subsection{Lensing and the SZ effect}\label{sec:cllens}

The thermal SZ effect is a
linear measure of the line-of-sight thermal energy content of a
cluster of galaxies. If the cluster is isothermal and the SZ effect
measures the line-of-sight electron density. Thus an SZ effect image
can be converted into a map of the line-of-sight integrated baryon
density. Since the baryon content of a cluster is dominated by the hot
gas in the atmosphere, rather than the baryons in galaxies, an SZ
effect image can be converted into a fairly accurate map of the total
baryon content of a cluster. This can be compared with the total mass
content of the cluster which underlies the gas density model used to fit
the SZ effect (or from some deprojected mass model underlying a
deprojection of the SZ data) to find baryonic mass fraction of the
cluster, and to see whether this changes across the cluster (as it
might if there is complicated 3-D structure: potentially this is a way
of recovering the 3-D structure of a cluster (Lee \atque Suto 2004).

More interestingly, the baryonic content map can be compared with the
total mass map derived from gravitational lensing to provide a
check of the physical description of the mass and gas in the cluster,
or to calculate the baryonic mass fraction using the SZ/X-ray derived
distance of the cluster rather than the Hubble constant
(Sec.~\ref{sec:lensing}).

For CL~0016+16, existing weak lensing data extend only to a radius
$\approx \theta_{\rm c}$ from the cluster centre (Smail \etal\ 1997),
so a comparison of the total cluster mass from lensing and implied by
the $\beta$ model is only possible in a cylinder of angular radius
$\theta_{\rm c}$ (corresponding to a linear scale of about $250 \ \rm
kpc$). The mass results are
\begin{equation} 
  M_{\rm tot} = \cases{ (2.0 \pm 0.1) \times 10^{14} \ \rm M_\odot 
                        & $\beta$ model, X-ray and SZ data \cr
                        (2.7 \pm 0.9) \times 10^{14} \ \rm M_\odot 
                        & weak lensing \cr
                      }
\end{equation}
and so are consistent although the lensing mass estimate has a
substantial error. By comparison, the total gas mass within the same
cylinder is 
\begin{equation} 
  M_{\rm gas} = (2.6 \pm 0.2) \times 10^{13} \ \rm M_\odot 
\end{equation}
implying a baryonic mass fraction in the cluster 
\begin{equation}
  f_{\rm b} = 0.13 \pm 0.04
\end{equation}
which is consistent with the baryonic mass fraction that would be
expected from a fair sample of matter in the Universe in the consensus
cosmology 
\begin{equation}
  {\Omega_{\rm b} \over \Omega_{\rm m}} = 0.12 \pm 0.02
\end{equation}
and shows no sign of evolution compared to nearby high-mass clusters
(Carlstrom \etal\  2002), even though the clusters for which this work
has been done do not at present constitute a well-defined sample.

\section{Next steps}

The studies of SZ effects to date have largely been performed on
relatively poorly-defined samples of clusters selected from ad-hoc
lists, often using criteria which could lead to those clusters being
unrepresentative of the total population, or even of clusters with
identical masses, because of accidental effects of orientation,
superposition, or transient dynamical state. Since the SZ effects are
linear probes of cluster properties such as total energy content, they
should provide excellent, well-defined, samples of clusters and
SZ-selected samples would be ideal for such tasks as constructing the
cluster-based Hubble diagram or examining the evolution of cluster
gravitational potentials.

A major focus of current SZ effect work is therefore to plan for, and
conduct, blind surveys of the sky to detect clusters solely through
their SZ effects. The full range of observing techniques can be
brought to bear on this task, but it is likely that radiometer and
bolometer arrays on single dishes will provide the fastest means of
performing surveys with arcminute angular resolution over
many square degrees of sky.

To date, most SZ effect work has concentrated on simple cluster
detection. However, with an increased number of detections, and the
prospect of a sample of SZ-selected clusters, it will soon become
important to follow up thse detections with higher angular-resolution
observations (which are likely to be interferometer-based) to study
the state of the gas within and around the 
selected clusters. Such observations might detect the kinematic
effects from rapidly-infalling filaments, or see substructure from the
accretion of filaments or subclusters. Further information can come
from polarization observations of clusters showing the brightest SZ
effects: measurements of both the radial and transverse
velocities of those clusters would provide the dynamical information
needed to make detailed tests of cluster formation models.

The requirements of these two programmes are somewhat different, and
so distinct types of telescope are needed for surveys and detailed
studies. Survey work is probably simplest using bolometer and
radiometer arrays (e.g., BOLOCAM or OCRA) although tailored
interferometers (e.g., AMI or AMiBA) can also be suitable. Detailed
cluster studies will need interferometers or bolometer arrays like
SCUBA-2 on large single dishes, since 10~arcsec or better angular
resolution is needed. Two examples of specially-designed instruments
are OCRA and AMiBA. 

\subsection{OCRA}

OCRA (Browne \etal~2000) was conceived as a 100-element cm-wave
radiometer array suitable for mounting at the prime or secondary focus
of a large single-dish telescope. A two-beam prototype (OCRA-p) is now
operational on the Torun 32-m antenna, with a larger array (OCRA-F) in
the final stages of completion. The sensitivity of the full OCRA is
such that it could map $100 \ \rm deg^2$ of sky to the confusion limit
in a few months, and should generate a sample of clusters of galaxies
at all redshifts where they have significant atmospheres, by virtue of
the redshift-independence of the brightness temperature change caused 
by SZ effects.

\subsection{AMiBA}

\begin{figure}
  \centering
  \includegraphics[width=0.7\linewidth]{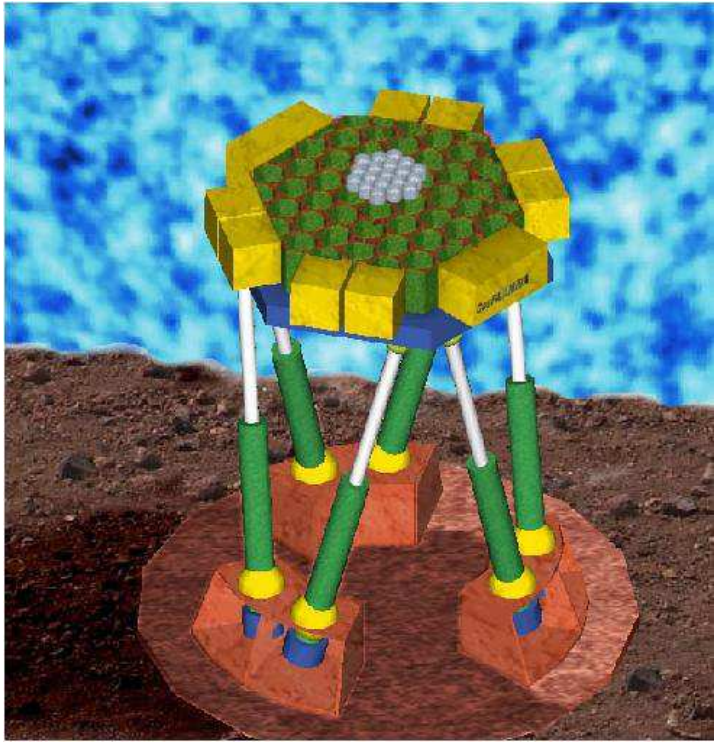}
  \caption{Artist's impression of the Array for Microwave Background
    Anisotropy (AMiBA).  The two dish sizes are approximately 0.3 and
    1.2 metres in diameter respectively.}
  \label{fig:amiba}
\end{figure}

AMiBA (Lo \etal~2001), shown in Fig.~\ref{fig:amiba}, is a dedicated MBR
interferometer designed for rapid surveys of arcminute-scale microwave
background structures, including the SZ effects. This ASIAA/NTU
project is now replacing the prototype that has been sited on Mauna
Loa for the past year with the operational facility. 

\begin{figure}
  \centering
  \includegraphics[width=\linewidth, clip=]{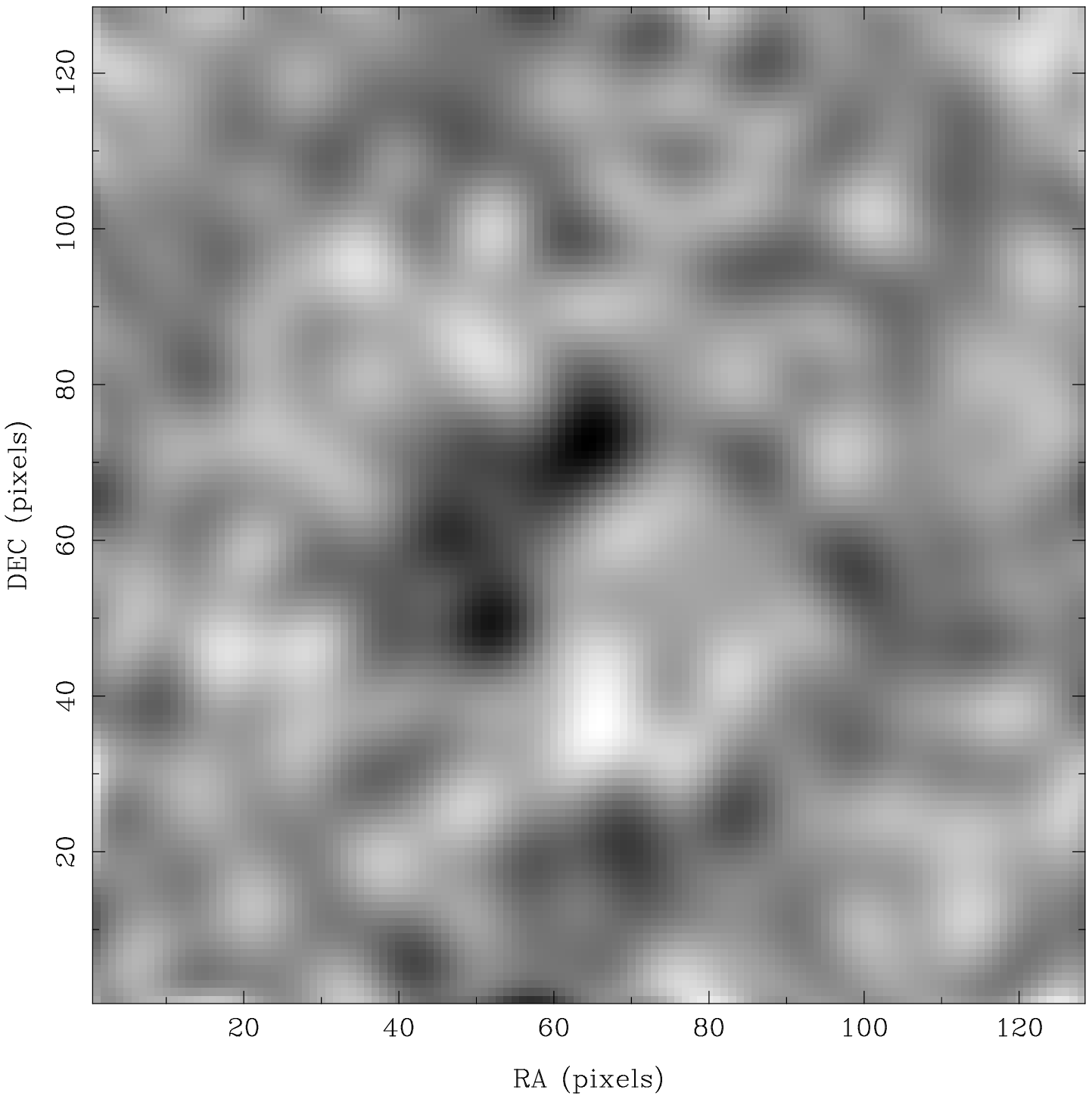}
  \caption{A simulated AMiBA field from the planned deep survey,
    showing two SZ detections.}
  \label{fig:amiba-map}
\end{figure}

AMiBA is planned to have 19~antennas of two sizes
(diameters 0.3 and 1.2 metres) and will offer a baseline range
$380\lambda < b < 1875\lambda$ at 95~GHz. The wide bandwidth at which
it operated ($\sim 20$~GHz), and dual-polarization operation, will
result in a sensitivity of about 1.3\,mJy in one hour, and a range of
surveys of different depths is planned to measure the distributions of
SZ effects in clusters at a range of redshifts. A simulated AMiBA
map, including two SZ detections, is shown in Fig.~\ref{fig:amiba-map}.

\subsection{The future: the ISZO}\label{sec:iszo}

Survey work requires the coverage of $100 \ \rm deg^2$ or more of sky
to a sensitivity of $30 \ \rm \mu K$ or better, and for greatest
effectiveness this should be possible in 1~year or less. Bolometer and
radiometer arrays, or interferometers, are all potentially capable of
making such surveys. 

However, it will be important to defeat the
confusion limit on such surveys imposed by primordial structure on
MBR, and to separate the thermal and kinematic SZ effects. This
requires a multi-wavelength survey, and one where the different
wavelength bands are well matched in angular resolution and
astrometric accuracy. The ideal facility for this type of work would
probably be a bolometer array, sited on a single large telescope, 
with the signals separated by dichroics, and sitting on an excellent
site. Experience of such facilities will come from BOLOCAM (Glenn
\etal~1998) and APEX (Schwan \etal~2003).

To study the structures of clusters, it will be necessary to achieve
sensitive sub-arcmin scale imaging over fields several arcmin in
size. While this is a natural ability of interferometers, it might
also be possible to achieve the necessary imaging quality using a
bolometer array. Again, multi-wavelength operation will be necessary
to separate the different SZ effects and to remove the contamination
from radio sources and star-forming galaxies, so that well-matched
bolometer arrays or scaled interferometer configurations will be
needed. An aim should be access to the velocity channel with a
sensitivity $\sim 100 \ \rm km \, s^{-1}$, which matches the noise
imposed by confusion with primordial structures, 
and some polarization capability, to allow a first investigation of
the polarization channel.
 
Many of these capabilities exist in the current generation of
instruments, but it is notable that these instruments are designed
at least roughly to match the sensitivity of the current generation
of X-ray telescopes. The next generation of X-ray satellites
(Con-X and XEUS) will have far higher sensitivity and better spectral
capabilities, and so will provide far superior images of clusters of
galaxies and of the thermal substructures within them. Now is
therefore the time to begin considering an SZ effect telescope that
could match these instruments in making detailed studies of cluster
substructure and evolution, by achieving at least an order of
magnitude improvement in sensitivity over currently-planned systems
such as AMI.

One possible design would see the construction
of scaled interferometers capable of operation at 30, 90, and 230~GHz
(to operate in good atmospheric windows). The scaled design is adopted
to permit similar areas of sky to be synthesized at each band: the use
of a single telescope design capable of covering all three bands would
lead to only small areas of the 230-GHz sky being accessible in a
single synthesis. Baselines would be from about $(300 - 10^4)\lambda$,
providing $\sim 20$~arcsec resolution and sensitivity to 
angular scales as large as $10$~arcmin. Individual antennas would be
$\sim 300 \, \lambda$ in size, to allow close packing at the smallest
antenna-antenna separations. About 20\% bandwidth at each band should
be possible, but the bands would need to be subdivided 
to avoid excessive bandwidth smearing. Full polarization
capability would allow the polarization channel to be explored (and
the sensitivity of the array would allow
some other polarization studies of large-scale structures). With $\sim
20$~antennas and non-uniform antenna separations, good instantaneous
sensitivity could be obtained over the full range of angular scales,
and with modern receivers the required sensitivity should be
achievable. 

An alternative design might be to use bolometer arrays at
110, 230, and 345~GHz on a telescope $\sim 50$~m in size (diffraction
limited at the longest wavelength). About 1000~elements would be
needed in each array, which is certainly feasible. This design could
have high instantaneous sensitivity, making rapid surveys possible.

Either system would be a powerful follow-up to Planck measurements of
the SZ effects of some thousands of clusters that we expect, and could
provide the deep surveys that would reach to $z > 1$ in the SZ
effects. However, both designs involve large-scale projects, and 
either would require an international team. This concept of an
International Sunyaev-Zel'dovich effect Observatory (ISZO) requires
considerable work to reach the level of a costed proposal, but it is
clear that such an observatory will be a requirement within 10~years
to take SZ effect studies to the next level of sophistication and
accuracy.

\subsection{Summary}

Twenty years ago the SZ effects were difficult to measure --- although
all-sky surveys were achieving excellent sensitivity to the
statistical distribution of MBR noise, it was difficult to measure an
arcminute-scale structure with a sensitivity of $100 \ \rm \mu K$, so
the existence of the SZ effect from any cluster was questionable.

Ten years ago the thermal SZ effect was regarded as a proven quantity,
with possibilities for providing new and interesting cluster and
cosmological information.

Today the SZ effects are seen as important windows into cluster
formation physics and cosmology, but they are not yet being used fully
because it remains difficult to detect any but the most luminous
clusters via their thermal SZ effects, and there is insufficient
cluster spectral information to make a statistical measure of cluster
kinematic SZ effects.

Over the next decade we will enter the era of precise SZ effect
measurements, and their routine use in cosmology and astrophysics. It
is now important to design the next generation of instruments, for the
new era of SZ effect observations, to match the coming generation of
high-sensitivity X-ray satellites.

\end{document}